\documentclass[journal]{IEEEtran}

\usepackage[T1]{fontenc}
\usepackage{caption}
\usepackage{graphicx}
\usepackage{epstopdf}
\usepackage{amsmath}
\usepackage{enumerate}
\usepackage[square,sort,numbers]{natbib}
\usepackage{algpseudocode}
\usepackage{subcaption}
\usepackage{float}
\usepackage{algorithm}
\usepackage{color}
\newfloat{algorithm}{t}{lop}

\let\ss= \scriptscriptstyle

\newcommand{\s}{\mathrm{s}}
\newcommand{\m}{\mathrm{m}}
\newcommand{\radX}[1]{r_\mathrm{#1}}
\newcommand{\radR}{\radX{r}}
\newcommand{\radD}{\radX{0}}
\newcommand{\rx}{\mathrm{r}}
\newcommand{\hit}{\mathrm{hit}}
\newcommand{\apriori}{\emph{a priori}}

\newcommand{\R}{R}
\newcommand{\RMC}{\textnormal{RMC}}
\newcommand{\APMC}{\textnormal{APMC}}
\begin{document}

\title{A Novel \emph{A Priori} Simulation Algorithm for Absorbing Receivers in Diffusion-Based Molecular Communication Systems}

\author{Yiran~Wang,~\IEEEmembership{Student~Member,~IEEE,}
        Adam~Noel,~\IEEEmembership{Member,~IEEE,}
        and~Nan~Yang,~\IEEEmembership{Member,~IEEE}\vspace{-2mm}

\thanks{The material in this paper was presented in part at the International Workshop on Molecular, Biological and Multiscale Communications, IEEE International Conference on Sensing, Communication and Networking (IEEE SECON 2018), in Hong Kong, China in June 2018~\cite{wang2018new}.}
\thanks{Y. Wang and N. Yang are with the Research School of Engineering, Australian National University, Canberra, ACT 2600, Australia (Emails: \{yiran.wang, nan.yang\}@anu.edu.au).}
\thanks{A. Noel is with the School of Engineering, University of Warwick, Coventry, CV4 7AL, UK (Email: adam.noel@warwick.ac.uk).}}

\markboth{Submitted to IEEE Transactions on NanoBioscience}%
{Wang \MakeLowercase{\textit{et al.}}: A Novel \emph{A Priori} Simulation Algorithm for Absorbing Receivers in Diffusion-Based Molecular Communication Systems}

\maketitle

\begin{abstract}
A novel \textit{a priori} Monte Carlo (APMC) algorithm is proposed to accurately simulate the molecules absorbed at spherical receiver(s) with low computational complexity in diffusion-based molecular communication (MC) systems. It is demonstrated that the APMC algorithm achieves high simulation efficiency since by using this algorithm, the fraction of molecules absorbed for a relatively large time step length precisely matches the analytical result. Therefore, the APMC algorithm overcomes the shortcoming of the existing refined Monte Carlo (RMC) algorithm which enables accurate simulation for a relatively small time step length only. Moreover, for the RMC algorithm, an expression is proposed to quickly predict the simulation accuracy as a function of the time step length and system parameters, which facilitates the choice of simulation time step for a given system. Furthermore, a likelihood threshold is proposed for both the RMC and APMC algorithms to significantly save computational complexity while causing an extremely small loss in accuracy.
\end{abstract}

\begin{IEEEkeywords}
Diffusion-based molecular communication, absorbing receivers, molecular communication simulation, Monte Carlo method.
\end{IEEEkeywords}

\IEEEpeerreviewmaketitle

\section{Introduction}\label{sec:Introduction}

\IEEEPARstart{M}{olecular} communication (MC) has emerged as an underpinning paradigm of exchanging and conveying information among nano-devices in very small dimensions or specific environments, such as water, tunnels, and human bodies \cite{nakanobook}. Unlike electromagnetic wave-enabled communication, MC delivers information based on chemical changes within the environment. In MC, transmitters first send out information-carrying molecules to propagate within the environment. Later, such molecules are captured by receivers to allow them to obtain the carried information. Biological examples of MC include chemotactic signaling, calcium signaling, and bacterial migration \cite{nakanobook}. Thus, MC offers the advantages of low energy consumption and potential for biocompatibility in liquid and gaseous media. Notably, the development of MC is envisioned to support transformational nano-applications, e.g., intra-body health monitoring, target drug delivery, and food and water quality monitoring \cite{nakanobook}.

Diffusion-based MC has been acknowledged as a simple but commonly adopted MC system within the nano-communication research community \cite{7405285}. In this system, information molecules propagate using kinetic energy only, which preserves a high energy efficiency. One of the major challenges in designing and analyzing a diffusion-based MC system is receiver modeling. The majority of existing MC studies have considered two types of receivers: passive receivers and active receivers \cite{7405285,nakano2012}. Passive receivers do not impose any impact on molecule propagation, while active receivers have some mechanism for molecules to react either within or at the surface of the receiver. The active receiver model is more general and is better representative of typical biological receivers.
This motivates us to investigate the properties of absorbing receivers,
which are a common ideal approximation for active receivers.

The notion of diffusion with absorption is a long-existing biological phenomenon that has been investigated in the literature, e.g., \cite{berg1993random,crank1956mathematics}. For example, \cite{berg1993random} investigated the diffusion into adsorbers, where the adsorbed molecules stop diffusion upon hitting the receiver, but later may desorb and resume diffusion. Specifically, \cite{berg1993random} examined the diffusion into a spherical adsorber, an ellipsoidal adsorber, and disk-like adsorber(s) in an infinite medium and obtained the diffusion currents for such adsorbers by solving Fick's equations. It is noted that some conclusions drawn for adsorbers can be applied to absorbers, due to the similarity between adsorption and absorption. Unlike \cite{berg1993random}, \cite{crank1956mathematics} discussed the absorption of molecules that can either diffuse through or react chemically with the receiver surface. Based on such studies, \cite{6807659,yilmaz2014simulation} considered a diffusion-based MC system with a single perfectly absorbing receiver within an unbounded three-dimensional (3D) environment. To be specific, \cite{6807659} provided the numerical results of the hitting rate of molecules at different times and the fraction of molecules absorbed for a given time, the analysis of which was performed in \cite{schulten2000lectures}, while \cite{yilmaz2014simulation} presented a simulation framework. Recently, \cite{7506290,ahmadzadeh2016comprehensive} evaluated the impact of a receiver with reversible adsorption on the performance of MC systems, where a molecule can be released back to the environment at some time after being captured by the receiver.

Apart from the theoretical analysis of MC systems, e.g., \cite{6807659,7506290,ahmadzadeh2016comprehensive}, the simulation of MC systems also serves as an effective means for performance evaluation. Against this background, a number of simulation frameworks, e.g., N3Sim \cite{llatser2014n3sim}, NanoNS \cite{gul2010nanons}, BiNS2 \cite{felicetti2013simulating}, and AcCoRD \cite{noel2017simulating}, have been developed to examine the behavior of information particles in various MC environments. The development of such frameworks is based on two common simulation approaches, the microscopic approach and the mesoscopic approach \cite{hellander2012coupled}. Considering a sub-region of the simulation environment, the microscopic approach treats the molecules within the region individually, while the mesoscopic approach treats them in aggregate as uniformly distributed throughout the region.
A widely recognized example of the microscopic approach in MC is the particle-based simulation that uses Brownian motion to characterize particle propagation. A detailed comparison between the microscopic and mesoscopic approaches and further information on how they have been implemented in MC frameworks were shown in \cite{noel2017simulating,farsad2016comprehensive}. In this paper, we focus on the particle-based microscopic approach due to its prevalence for the physical simulation of an MC environment.

It is worthwhile noting that the existing microscopic algorithms incur high computational complexity in the simulation of the fraction of molecules absorbed in MC systems with absorbing receivers. This high complexity leads to a long simulation run time since the existing algorithms require a very small time step length to accurately model the absorption. To tackle this issue, we develop a new particle-based microscopic algorithm in this work. This algorithm aims to reduce the computational complexity for the simulation of an irreversible perfectly absorbing receiver in a 3D MC system where every molecule that arrives at the receiver is absorbed.

In the simulation procedure of the particle-based microscopic algorithm for diffusion-based MC systems, molecules are moved by adding Gaussian random variables (RVs) to their $x$-, $y$-, and $z$-coordinates at the end of every simulation time step. This makes the coordinates discrete functions of time, although in reality the movements of molecules undergoing Brownian motion are continuous over both time and space. In practice, molecules may actually diffuse into an absorbing receiver between two time steps. To determine this absorption, some existing simulation algorithms simply compare the observed coordinates of molecules after diffusing with the coordinates of the receiver \cite{yilmaz2014simulation,7506290,llatser2014n3sim}. As a consequence, the molecules that hit a receiver between time steps cannot be considered as ``absorbed'' by using these simulation algorithms. We name the probability that a molecule crosses the RX's boundary between time steps as the \emph{intra-step absorption probability}. Recently, \cite{noel2017simulating} and \cite{arifler2017} investigated the possibility of intra-step absorption. Specifically, \cite{noel2017simulating} declared a molecule as ``absorbed'' if its straight-line trajectory within a time step crossed an absorbing surface. Alternatively, \cite{arifler2017} approximated the intra-step absorption probability for spherical receiver boundaries using the equation for flat planar receiver boundaries given by \cite[Eq.~(10)]{andrews2004stochastic}. This approximation was referred to as the refined Monte Carlo (RMC) algorithm.

Our paper focuses on improving the accuracy and reducing the computational cost of simulating absorbing receivers. We first run simulations using the RMC algorithm and measure the change in accuracy 
as the time step length increases. Then, we fit the results to an expression to estimate the accuracy given the diffusion coefficient $D$, time step length $\Delta{}t$, RX's radius $\radR$, and transmitter-to-receiver distance $\radD$. After recognizing the poor accuracy of the RMC algorithm when $\Delta{}t$ increases, we propose a new method for simulations of absorbing receiver(s) with a large time step length, namely, the $\apriori$ Monte Carlo (APMC) algorithm. The APMC algorithm uses the $\apriori$ probability of a molecule being absorbed to decide whether it is absorbed in the current simulation time step. If the molecule is determined as ``absorbed'' using the $\apriori$ probability, then we omit the diffusion step. We show that this algorithm achieves very high accuracy when the diffusion step length is large relative to the size of the receiver. Despite that the RMC algorithm performs accurately for small time step size, we observe from MATLAB that the simulation run time of the RMC algorithm is higher than that of the APMC algorithm for small time step size. Furthermore, we identify that a major contributor to the computational complexity of the RMC and APMC algorithms is the generation of uniform RVs when assessing intra-step absorptions. Thus, we propose a likelihood threshold for both algorithms to reduce the computational cost caused by the generation of excessive RVs.

Our contributions extend our preliminary work in \cite{wang2018new} and are summarized as follows:
\begin{enumerate}
\item As in \cite{wang2018new}, we present a new algorithm, i.e., the APMC algorithm, for MC simulation with a relatively large time step length. The advantages of the APMC algorithm are:
    \begin{enumerate}
    \item For the case of a single perfectly absorbing receiver, we show that by using the APMC algorithm, the fraction of molecules absorbed precisely matches the corresponding analytical result when $\sqrt{D \Delta t}/\radR$ is relatively large.
    \item For the case of two perfectly absorbing receivers, we show that by using the APMC algorithm, the fraction of molecules absorbed approaches the asymptotic analytical value when time grows large.
    \end{enumerate}
    We note that the advantages of the APMC algorithm in both cases cannot be achieved by the existing algorithms.
  \item We propose polynomial fitting expressions to predict the accuracy of the RMC algorithm introduced in \cite{arifler2017} for the case of a single perfectly absorbing receiver. Aided by numerical results, we show that the third order polynomial fitting expression is the most accurate one for accuracy prediction. This allows us to use this expression to characterize the accuracy of the simulation without running it.
  \item We investigate the computational complexity of both the RMC algorithm and the APMC algorithm. Specifically, we compare their MATLAB run times and explore the trade-off between simulation accuracy and computational complexity for both algorithms, based on which we propose a likelihood threshold for the absorption probability to save computation time. Molecule absorption for the RMC algorithm and the APMC algorithm is possible if and only if the calculated absorption likelihood in the simulation is higher than the likelihood threshold. Aided by numerical results, we show that applying a likelihood threshold can save as much as 20\% of the total number of generated RVs with an extremely small loss in simulation accuracy. 
\end{enumerate}

Comparing to our preliminary work [1], which only presented the received signals of the SMC, RMC and APMC algorithms for the system with a single absorbing receiver, this paper additionally presents the received signals for multi-receiver configurations, proposes prediction expressions for the RMC algorithm, and investigates the computational complexity of the RMC and APMC algorithms. Furthermore, the majority of the numerical results and discussions shown in this paper are not included in [1]. 

The rest of the paper is organized as follows. The system model of interests is described in Section \ref{sec:System_Model}. Existing simulation algorithms, our proposed APMC algorithm, and the likelihood threshold are presented in Section \ref{sec:Simulation_Algorithms}. Numerical results and discussions are provided in Section \ref{sec:Numerical_Results}. In Section \ref{sec:Conclusion}, we present our conclusions.

\section{System Model}\label{sec:System_Model}

\begin{figure}[t]
    \centering
    \includegraphics[width=0.4\textwidth]{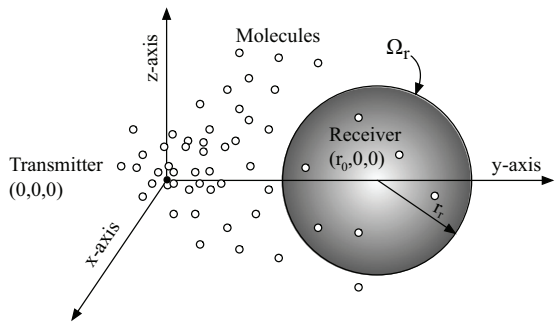}
    \caption{Illustration of our system model. The TX is a point transmitter located at $(0,0,0)$, the RX is a spherical irreversible perfectly absorbing receiver located at $(\radD,0,0)$ with $\radR$ being the radius and $\Omega_\textnormal{r}$ being the RX's perfectly absorbing boundary. Molecules propagate in the environment according to Brownian motion.\vspace{-1em}}\label{MC system illustration}
\end{figure}

\begin{figure}[t]
    \centering
    \includegraphics[width=0.44\textwidth]{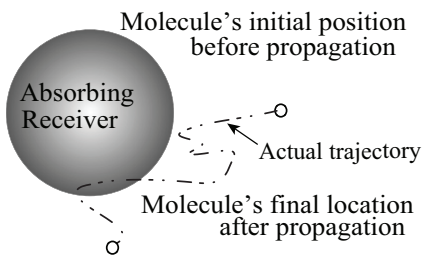}
    \caption{Illustration of the intra-step molecule movement. There is a possibility that a molecule crosses an absorbing boundary within one time step, even if its initial and final positions during that time step are both outside the absorbing receiver.\vspace{-1em}}\label{intra-step absorption}
\end{figure}

We consider a diffusion-based MC system within a 3D space, as depicted in Fig.~\ref{MC system illustration}. In this system, a point transmitter (TX) is located at the origin of the space and a spherical perfectly absorbing receiver (RX) is centered at location $(\radD,0,0)$. We denote $\radR$ as the RX's radius and $\Omega_{\rx}$ as the RX's boundary. At the beginning of a transmission process, the TX instantaneously releases $N$ molecules. We assume that the molecules are small enough to be considered as points. Once released, molecules diffuse in the environment according to Brownian motion until hitting the RX's boundary. We denote $N_{\hit}\left(\Omega_{\rx},t|\radD\right)$ as the number of molecules released from the TX at time $t_\textnormal{0} = 0\,\s$ and absorbed by the RX by time $t$. As per \cite[Eq.~(3.116)]{schulten2000lectures}, we express $N_{\hit}(\Omega_{\rx},t|\radD)$ as
\begin{equation}\label{yilmazequation}
N_{\hit}\left(\Omega_{\rx},t|\radD\right)=
\frac{N\radR}{\radD}\textnormal{erfc}\left(\frac{\radD-\radR}{\sqrt{4Dt}}\right),
\end{equation}
where $D$ is the diffusion coefficient and $\textnormal{erfc}\left(\cdot\right)$ is the complementary error function. $D$ describes the proportionality constant between the flux due to molecular diffusion and the gradient in the concentration of molecules.

The majority of the existing simulation algorithms for absorbing RXs did not consider the possibility of \emph{intra-step molecule absorption} \cite{arifler2017}. This absorption is depicted in Fig.~\ref{intra-step absorption}, which shows that the actual trajectory of a molecule may cross the RX's boundary during one simulation time step, even if its initial position at the beginning of the time step and its final position at the end of the same time step are both outside the absorbing RX. If this crossing occurs, the molecule is absorbed by the RX in practice. When the possibility of this absorption is ignored, the number of absorbed molecules is underestimated, thus deteriorating the accuracy of simulation. In this work, we refer to the probability that a molecule is absorbed during a simulation time step as the \emph{intra-step absorption probability}. For MC systems, the discussion of this probability was limited. Two of the papers that considered the \emph{intra-step absorption probability} are \cite{noel2017simulating,arifler2017}. In \cite{noel2017simulating}, the absorption of a molecule was determined by whether its straight-line trajectory has crossed the absorbing surface. In \cite{arifler2017}, the intra-step absorption probability of a perfectly absorbing RX with a spherical boundary was approximated as that of a perfectly absorbing RX with an infinite \emph{planar} boundary, given by \cite[Eq. (10)]{andrews2004stochastic}
\begin{align}\label{Andrews2004}
{\textrm{Pr}}_{\ss\RMC}=\exp{\left(-\frac{l_\textrm{i}l_\textrm{f}}{D\Delta t}\right)},
\end{align}
where $l_\textrm{i}$ is the initial distance of a molecule from the absorbing boundary at the beginning of a time step, $l_\textrm{f}$ is the final distance of a molecule from the absorbing boundary at the end of the same time step, and $\Delta t$ is the time step length.

Apart from the aforementioned MC system which contains only one absorbing spherical RX, we also consider a system containing two identical absorbing spherical RXs. In this system, TX stays at the origin of the space and two absorbing RXs are located on opposite sides of and equidistant from the TX. Focusing on this two-RX system, \cite{sano1981solutions} evaluated the fraction of molecules absorbed, which is defined as the ratio between the number of molecules absorbed at the RX and the total number of molecules released. As the time elapsed goes to infinity, the asymptotic fraction of molecules absorbed by each RX is given by \cite{sano1981solutions}
\begin{align}\label{solution}
&\textnormal{Pr}_{t\to\infty}=\sqrt{2\left(\cosh\mu -\cos\eta\right)}\notag\\
&\hspace{2mm}\times\sum_{n=0}^{\infty}\mathrm{e}^{-\left(n+\frac{1}{2}\right)\mu_1}
\frac{\sinh{\left(n+\frac{1}{2}\right)\left(\mu-\mu_2\right)}}
{\sinh{\left(n+\frac{1}{2}\right)\left(\mu_1-\mu_2\right)}}P_n\left(\cos(\eta)\right),
\end{align}
where $\left(\mu, \eta, \phi\right)$ are the bispherical coordinates corresponding to $(0, 0, 0)$ in the natural coordinate system, $\mu_1 = \cosh^{-1}\left(\radD/\radR\right)$, $\mu_2 = -\cosh^{-1}\left(\radD/\radR\right)$, $\radR$ is the RX's radius, $\radD$ is the distance from the center of one RX to the TX, and $P_n$ is the $n$th-degree Legendre polynomial. The two-RX system allows us to show that our proposed simulation algorithm can be applied to not only a single-RX system but a multi-RX system, which will be demonstrated in Section \ref{sec:Numerical_Results}.

\section{Simulation Algorithms}\label{sec:Simulation_Algorithms}
\begin{figure}[t]
    \centering
    \includegraphics[width=0.39\textwidth]{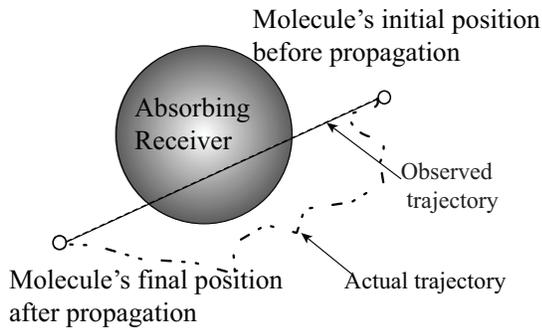}
    \caption{An example which shows the inaccuracy of the determination criterion in \cite{noel2017simulating}. In this example, the line segment from the molecule's initial position to its final position crosses the RX's surface, indicating that this molecule is absorbed by the RX as per the determination criterion in \cite{noel2017simulating}. However, the molecule's actual trajectory indicates that this molecule is not absorbed.}
    \label{line trajectory absorption}
\end{figure}
\begin{figure}[t]
    \centering
    \includegraphics[width=0.37\textwidth]{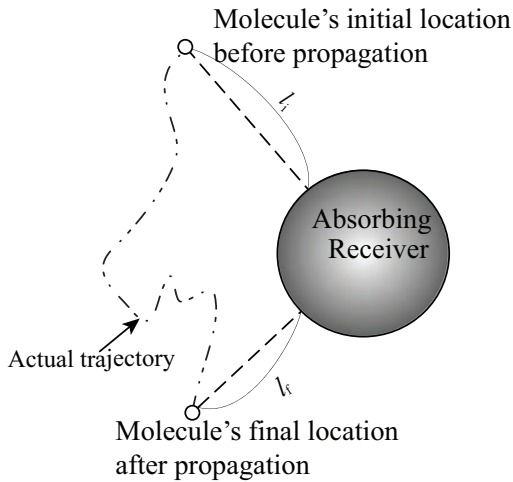}
    \caption{Illustration of the distances used in the intra-step absorption probability calculation in the RMC algorithm \cite{arifler2017}.}\label{cross}
\end{figure}
In this section, we first describe the common structure of the existing simulation algorithms for absorbing RXs. Then we propose polynomial functions to predict the accuracy of the RMC algorithm. After that, we present our proposed APMC algorithm and determine the likelihood threshold to be applied to the APMC and RMC algorithms. The algorithms discussed in this paper are summarized in Table~\ref{tablealgorithm}.
\begin{table}[t]
\caption{Comparison of Simulation Algorithms}\label{tablealgorithm}
\renewcommand{\arraystretch}{1.2}
\centering
\begin{tabular}{c||c|l}
\hline
$\textbf{Algorithm}$ & Diffusion first? & Intra-step absorption? \\\hline
\hline
SMC \cite{yilmaz2014simulation,crank1956mathematics}& Yes& No \\\hline
RMC \cite{arifler2017} & Yes& Yes. Absorption probability is given\\
& &by \eqref{Andrews2004}.\\\hline
AcCoRD \cite{noel2017simulating} &Yes& Yes. Absorption occurs when the\\
& &molecule trajectory crosses boundary.\\\hline
APMC &No &No. Molecules are absorbed before\\
& &being diffused.\\\hline
\end{tabular}
\end{table}

\subsection{Existing Simulation Algorithms}\label{sec:Simulation_Algorithms_Existing}

By observing the existing simulation algorithms for microscopic molecule absorption (such as those in \cite{yilmaz2014simulation,noel2017simulating,arifler2017}), we identify that they follow a common structure. This structure is presented in~\textbf{Algorithm~\ref{globalalgorithm}}. For simulating molecules that follow Brownian motion, $\Delta t$ needs to be carefully chosen such that the diffusion step length is relatively \emph{small} compared to the distances that define reflective or absorbing boundaries \cite{andrews2004stochastic}. We note that the ratio of the step length to the distance affects the performance of the simulation algorithms, as will be demonstrated in Section \ref{sec:Numerical_Results}.

\begin{algorithm}
\caption{A common structure of simulation algorithms for a single absorbing RX} \label{globalalgorithm}
\begin{algorithmic}[1]
\State Determine the end time of simulation.
\ForAll {simulation time steps}
\If {$t=0$}
\State Add $N$ molecules to environment.
\EndIf
\State Scan all \emph{not-yet-absorbed} molecules, i.e., molecules which are not absorbed by the RX.
\ForAll {\emph{not-yet-absorbed} molecules}
\State Propagate each molecule for one step according to Brownian motion.\label{alg:normalRV1}
\State Determine if the molecule is absorbed. \label{alg:absorption}
\EndFor
\EndFor
\end{algorithmic}
\end{algorithm}
\begin{algorithm}
\caption{Molecule absorption determination in \cite{arifler2017}}\label{algorithmRMC}
\begin{algorithmic}[1]
\If {The molecule's distance to $(\radD,0,0)$ is smaller than $\radR$,}
\State The molecule is absorbed.
\Else
\State Calculate the intra-step absorption probability, $\textnormal{Pr}_{\ss\RMC}$, using \eqref{Andrews2004}.
\State Generate a uniform RV $u$.
\If {$\textnormal{Pr}_{\ss\RMC}\geq{}u$}\label{alg:rmcabsorptionline}
\State The molecule is absorbed.
\EndIf
\EndIf
\end{algorithmic}
\end{algorithm}

We clarify that each algorithm has its own criterion for determining whether or not a molecule is absorbed by the RX, as given in Line~\ref{alg:absorption} of \textbf{Algorithm~\ref{globalalgorithm}}. Such determination criteria for algorithms in \cite{yilmaz2014simulation,noel2017simulating,arifler2017} are summarized as follows:
\begin{itemize}
\item
As per the determination criterion in \cite{yilmaz2014simulation}, the molecules being observed inside the RX at the end of a time step are absorbed. This criterion is referred to by \cite{arifler2017} as the simplistic Monte Carlo (SMC) algorithm. We note that the performance of the SMC algorithm is inaccurate, unless the time step length is \emph{very} small. This is because the SMC algorithm ignores the possibility of intra-step absorption and thus, underestimates the number of absorbed molecules.
\item
As per the determination criterion in \cite{noel2017simulating}, a molecule is absorbed if the line segment from its initial position to its final position crosses the RX's boundary. However, we note that the line segment crossing the RX's surface is neither sufficient nor necessary to correctly detect intra-step absorption. For example, in Fig.~\ref{intra-step absorption} the molecule absorption that actually occurs cannot be detected by the criterion in \cite{noel2017simulating}. In another case, shown in Fig.~\ref{line trajectory absorption}, molecule absorption is determined by the criterion in \cite{noel2017simulating} but does not actually occur. The accuracy of the algorithm in \cite{noel2017simulating} is better than but still comparable to the SMC algorithm.
\item
As per the determination criterion in \cite{arifler2017}, referred to as the RMC algorithm and described in~\textbf{Algorithm~\ref{algorithmRMC}}, \eqref{Andrews2004} is used to calculate the intra-step absorption probability of an RX with a perfectly absorbing spherical boundary. As depicted in Fig.~\ref{cross}, in the RMC algorithm, $l_\textnormal{i}$ denotes the shortest distance between the molecule's initial position and the RX's boundary and $l_\textnormal{f}$ denotes the shortest distance between the molecule's final position and the RX's boundary. As shown in \cite{arifler2017}, the accuracy of the SMC algorithm is comparable to that of the RMC algorithm when $\Delta t$ is relatively small. When $\Delta t$ increases while other parameters remain the same, the accuracy of the RMC algorithm becomes measurably higher than that of the SMC algorithm. However, in addition to the results presented in \cite{arifler2017}, we run simulations of the SMC and RMC algorithms and observe that the simulated fraction of absorbed molecules of the RMC algorithm deviates from the analytical one given by \eqref{Andrews2004} when the root mean square (RMS) of the diffusion step length, $\sqrt{2D\Delta t}$, is relatively larger than $\radR$. We will show the impact of $\Delta t$ on the accuracy of the RMC algorithm in Section \ref{sec:Numerical_Results}.
\end{itemize}

\subsection{Performance Prediction of RMC Algorithm}\label{sec:Simulation_Algorithms_Performance}

In this subsection, we propose a rule-of-thumb expression to predict the accuracy of the RMC algorithm when the parameters of the MC system are within specified ranges. Specifically, our aim is to make a rough but fast prediction of the algorithm's accuracy for given parameters, without resorting to simulations. To collect necessary data for accuracy prediction, we adopt the following procedure (the intermediate plots mentioned in the procedure are not presented in this paper due to page limit):
\begin{enumerate}
\item For the parameters listed in Table~\ref{tableaccuracy}, we run simulations using the RMC algorithm for one time step. We note that the fraction of molecules absorbed at every current time step is based on the results generated in the previous time step. In order to eliminate the impact of results from previous time steps, we choose to test for the first simulation time step only, which means that we calculate with only two samples, one at the very beginning of the transmission, and another at the end of the first time step. Simulation is repeated for $N$ molecules and the result obtained from the simulation of each molecule is called a realization. For each set of $D$, $\Delta t$, and $\radD$, we plot $\R^2$ versus $\radR$, where the measure of accuracy, $\R^2$, is calculated according to
    \begin{equation}\label{R}
    \R^2=1-\frac{\sum_{i=1}^2\left({\textnormal{Pr}}_{\hit}\left(i-1\right)-
    {{\textnormal{Pr}}}_{\textnormal{sim}}\left(i-1\right)\right)^2}
    {\sum_{i=1}^2\left({{\textnormal{Pr}}}_{\textnormal{sim}}\left(i-1\right)-
    \overline{{{\textnormal{Pr}}}}_{\textnormal{sim}}\right)^2},
    \end{equation}
    where
    \begin{equation}\label{fraction}
    {\textnormal{Pr}}_{\hit}(i-1)=N_{\hit}\left(\Omega_{\rx},(i-1)\Delta t|\radD\right)/N,
    \end{equation}
    is the analytical fraction of absorbed molecules at time $(i-1)\Delta t$, as obtained from \eqref{yilmazequation}. We define ${\textnormal{Pr}}_{\textnormal{sim}}(i-1)$ as the simulated fraction of absorbed molecules by the RMC algorithm at time $(i-1)\Delta t$, which is obtained by averaging the total number of absorbed molecules out of all realizations at $(i-1)\Delta t$ over $N$, and refer to it as the $i$th simulated sample. Additionally, we define $\overline{{\textnormal{Pr}}}_{\textnormal{sim}}$ as the mean of all simulated samples. 
\item We plot reference lines that indicate different values of $\R^2$ in the same figure.
\item We observe the values of $\radD$, $\radR$, $D$, and $\Delta t$ that achieve the selected values of $\R^2$.
\item For each selected value of $\R^2$, we plot a 3D figure where the $z$-coordinate is the recorded value of $\radR$, the $x$-coordinate is the recorded value of $D\Delta t$, and the $y$-coordinate is the recorded value of $\radD$.
\end{enumerate}

By observing the obtained results, we find that the accuracy of the RMC algorithm decreases when $D \Delta t$ increases, $\radD$ increases or $\radR$ decreases. To predict the accuracy, we define a dimensionless variable  $\kappa$ that is calculated from the parameters of the MC system. We assume linear dependency between the variables, i.e., $\kappa$ and $\radR$, for the sake of a quick and simple calculation. 
Using the curve fitting tool in MATLAB, we fit the 3D plots to equations of $\kappa$. Here, we define $\kappa$ as
\begin{equation}\label{estimationeq}
\kappa=\radR\left(\radD D\Delta t\right)^{-\frac{1}{3}},
\end{equation}
where the exponent of $-\frac{1}{3}$ is chosen such that the right hand side is dimensionless.
We next fit $\R^2$ to a polynomial function of $\kappa$. We test the first, second, and third order polynomial fits\footnote{We do not test higher order fits in this paper because we seek an easy-to-compute expression.}, which are given by
\begin{equation}\label{1storder}
R^2\approx\frac{101\kappa+47}{100},
\end{equation}
\begin{equation}\label{2storder}
R^2\approx\frac{-372\kappa^2+392\kappa-3}{100},
\end{equation}
and
\begin{equation}\label{3rdorder}
R^2\approx\frac{979\kappa^3-1523\kappa^2+813\kappa-51}{100},
\end{equation}
respectively. Given that $R^{2}$ is the accuracy of the RMC algorithm, which implies $0\leq\R^2\leq1$, we impose $0.0726\leq\kappa\leq0.612$ for (\ref{3rdorder}). If $\kappa > 0.612$, then the RMC algorithm is predicted to have asymptotically high accuracy. If $\kappa<0.0726$, then the RMC is predicted to have asymptotically low accuracy. We then arrive at
\begin{equation}\label{performance2}
R^2\approx
\begin{cases}
0, &\text{if}~\kappa<0.0726,\\
\frac{979\kappa^3-1523\kappa^2+813\kappa-51}{100}, & \text{if}~0.0726\leq\kappa\leq0.612, \\
1, &\text{otherwise}.
\end{cases}
\end{equation}
We will demonstrate in Section \ref{sec:Numerical_Results} that the third order polynomial fit is the best match among (\ref{1storder}), (\ref{2storder}), and (\ref{performance2}). The accuracy of the three polynomial fits will be shown in Fig.~\ref{estimation} and the corresponding measured RMS error (RMSE) will be shown in Table \ref{tableRMSE}.

\begin{table}[!t]
\caption{Range of MC System Parameters for Performance Prediction}\label{tableaccuracy}
\renewcommand{\arraystretch}{1.2}
\centering
\begin{tabular}{c||cc}
\hline
$\textbf{Parameter}$ & $\textbf{Notation and Range}$\\\hline
\hline
TX-RX distance & $20\,\mu\m\leq\radD\leq100\,\mu\m$ \\\hline
Radius of RX & $0\,\mu\m<\radR\leq\radD\,\mu\m$ \\\hline
RMS of diffusion step length & $40\,\mu\m\leq\sqrt{2D\Delta t}\leq\sqrt{20000}\,\mu\m$ \\\hline
Accuracy & $\R^{2}\in\{0.6,0.65,0.7,0.75,0.8,$\\ & $~~~~~~0.85,0.9,0.95,0.99\}$ \\\hline
\end{tabular}
\end{table}

\subsection{New A Priori Monte Carlo Algorithm}\label{sec:Simulation_Algorithms_New}

In this subsection, we propose a new simulation algorithm for approximating the fraction of molecules absorbed at a perfectly absorbing RX when $\sqrt{D\Delta t}/\radR$ is larger than that considered for the RMC algorithm in \cite{arifler2017}. We refer to the newly proposed algorithm as the APMC algorithm.

The procedure of the APMC algorithm is to first calculate, \emph{before} the $j$th molecule diffuses, the probability that this molecule will be absorbed in the \emph{current} time step. This probability depends on the distance between this molecule and the center of the RX, $d_{j}$, and the time step length, $\Delta{t}$. Specifically, the probability that this molecule will be absorbed in the \emph{current} time step is calculated as
\begin{equation}\label{analysisequationtimestep}
{\textrm{Pr}}_{\ss\APMC}=\frac{\radR}{d_j}\textrm{erfc}\left(\frac{d_j-\radR}{\sqrt{4D\Delta t}}\right),
\end{equation}
which is obtained by scaling \eqref{yilmazequation} by $N$, replacing the total simulation time $t$ with $\Delta t$, and replacing $\radD$ with $d_j$. In \eqref{analysisequationtimestep}, ${\textrm{Pr}}_{\ss\APMC}$ denotes the fraction of absorbed molecules released from a location $d_j$ away from the RX at time $t_\textnormal{0}=0\,\s$ and absorbed by the RX by time $\Delta t$. We note that \eqref{analysisequationtimestep} is calculated repeatedly in every time step for each of the free molecules with the molecule's current updated location when it diffuses. Then the molecule absorption is determined by generating a uniform RV $u$, where $0\leq{}u\leq1$, and comparing its value with the probability obtained by \eqref{analysisequationtimestep}. A molecule is marked as ``absorbed'' if $u\leq{\textrm{Pr}}_{\ss\APMC}$. After determining the molecules absorbed, each not-yet-absorbed molecule is propagated according to Brownian motion.  If any propagated molecule is inside the RX's boundary at the end of the current time step, we \emph{revert} the diffusion of this molecule and let it propagate again, until this molecule diffuses to a location outside the RX. This is because if a molecule propagates to a location inside the RX, it contradicts the preconditioning that the molecule is not absorbed. The APMC algorithm is detailed in \textbf{Algorithm \ref{algorithmapriori}}.

\begin{algorithm}
\caption{The APMC algorithm for molecule absorption}\label{algorithmapriori}
\begin{algorithmic}[1]
\State Determine the end time of simulation.
\ForAll {simulation time steps}
\If {$t = 0$}
\State Release $N$ molecules into environment.
\EndIf
\State Scan all \emph{not-yet-absorbed} molecules.
\ForAll {\emph{not-yet-absorbed} molecules}
\State Calculate the distance between the $j$th molecule to $(\radD,0,0)$, denote by $d_j$.
\State Calculate the absorbed probability $\textnormal{Pr}_{\ss\APMC}$ for each \emph{not-yet-absorbed} molecule using \eqref{analysisequationtimestep} with $\radR$, $d_j$, $D$, and $\Delta t$.
\If { $\textnormal{Pr}_{\ss\APMC}\geq u$} \label{alg:aprioriabsorptionline}
    \State The molecule is absorbed.
    \EndIf
    \EndFor
\ForAll {\emph{not-yet-absorbed} molecules}
\State Propagate the molecule for one step. \label{alg:APMCRV1}
\While {the molecule's distance to $(\radD,0,0)\leq\radR$}
    \State Revert the movement of this molecule to the location before propagation.
    \State Propagate this molecule again. \label{alg:APMCRV2}
\EndWhile
\EndFor
\EndFor
\end{algorithmic}
\end{algorithm}

\subsection{Likelihood Threshold for Simulation Complexity Reduction}\label{sec:Simulation_Algorithms_Rejection}

In this subsection, we propose a likelihood threshold to reduce the computational complexity for the RMC algorithm and the APMC algorithm. For both algorithms, the computational complexity consists of three parts and can be written as
\begin{equation}
\mathcal{C}_{\textnormal{sim}} = \mathcal{O}\left(\frac{t_\textnormal{end}}{\Delta t} \left(c_{\textnormal{diffuse}}+c_{\textnormal{absorb}}+c_{\textnormal{locate}}\right)\right),
\end{equation}
where $c_{\textnormal{diffuse}}$ denotes the complexity of diffusing all molecules, $c_{\textnormal{absorb}}$ denotes the complexity of determining the molecules to be absorbed and absorbing these molecules, and $c_{\textnormal{locate}}$ denotes the complexity of determining whether a molecule is inside the spherical RX. 
We observe from MATLAB run time profiles that the time for generating uniform RVs alone can take up around $10\%$ of the total simulation run time of both algorithms and that an increase in the number of generated RVs is closely related to an increase in the total simulation run time for both algorithms. Therefore, we use the number of generated RVs to characterize the computational complexity of both algorithms. 

There are two types of RVs for the APMC algorithm and the RMC algorithm, namely, the uniform RVs used for testing the molecule absorption and the Gaussian RVs used for propagating molecules according to Brownian motion. Thus, we denote $\mathcal{N}_\textrm{u}$ and $\mathcal{N}_\textrm{g}$ as the number of uniform RVs and the number of Gaussian RVs, respectively. For each determination of molecule absorption, as shown in Line \ref{alg:rmcabsorptionline} in \textbf{Algorithm \ref{algorithmRMC}} and Line \ref{alg:aprioriabsorptionline} in \textbf{Algorithm \ref{algorithmapriori}}, we add $1$ to $\mathcal{N}_\textrm{u}$. For each molecule propagation, as shown in Line \ref{alg:normalRV1} in \textbf{Algorithm \ref{globalalgorithm}} and Lines \ref{alg:APMCRV1} and \ref{alg:APMCRV2} in \textbf{Algorithm \ref{algorithmapriori}}, we add $3$ to $\mathcal{N}_\textrm{g}$ since there are three Gaussian RVs added to the $x$-, $y$-, and $z$-coordinates of a molecule. Here, we adopt the commonly used Box-Muller transform \cite{marsaglia} to convert the number of generated Gaussian RVs to an equivalent number of generated uniform RVs. According to \cite{marsaglia}, the generation of a pair of Gaussian RVs requires $4/\pi$ pairs of uniform RVs. Thus, we use $4/\pi$ as a conversion factor and then the equivalent number of totally generated uniform RVs, which characterizes the computational complexity of simulation, is given by $\mathcal{N}_{\textrm{total}}=\mathcal{N}_\textrm{u}+\left(4/\pi\right)\mathcal{N}_\textrm{g}$.

Based on the aforementioned computational complexity characterization, we now propose a likelihood threshold, $\xi$, to reduce the computational complexity. With this likelihood threshold, molecule absorption for the RMC algorithm is possible if and only if $\textrm{Pr}_{\ss\RMC}\geq\xi$, while molecule absorption for the APMC algorithm is possible if and only if $\textrm{Pr}_{\ss\APMC}\geq\xi$. This can significantly reduce $\mathcal{N}_\textrm{u}$ in simulation. Although the accuracy of the RMC and APMC algorithms may slightly decrease when $\xi$ applies, an appropriate value of $\xi$ can provide a good trade-off between simulation accuracy and computational complexity. Such trade-off of the RMC and APMC algorithms will be illustrated in Section \ref{sec:Numerical_Results}.

\section{Numerical Results}\label{sec:Numerical_Results}

In this section, we compare the fraction of molecules absorbed determined by the SMC algorithm, the RMC algorithm, and the APMC algorithm, with the aid of particle-based simulations, to show the benefits of our APMC algorithm. We also investigate the accuracy of the prediction expression obtained from empirical simulations in Section~\ref{sec:Simulation_Algorithms_Performance} by examining the performance of the RMC algorithm. Furthermore, we compare simulation run times of the RMC and APMC algorithms and explore the trade-off between simulation accuracy and computational complexity of algorithms. The algorithms are implemented in MATLAB.

Throughout this section, we denote $M$ as the number of time-varying samples, and hence $M-1$ as the number of time steps. Unless otherwise stated, we adopt the diffusion coefficient of $D = 10^{-9}\,{\m^2}/{\s}$, the TX-RX distance of $\radD = 50\,\mu\m$, and the number of molecules released of $N = 10^6$.

\subsection{A Single Absorbing Receiver}\label{sec:Numerical_Results_Single}

\begin{figure}[!t]
\centering
    \begin{subfigure}[b]{0.24\textwidth}
    \includegraphics[width=\textwidth,height=2in]{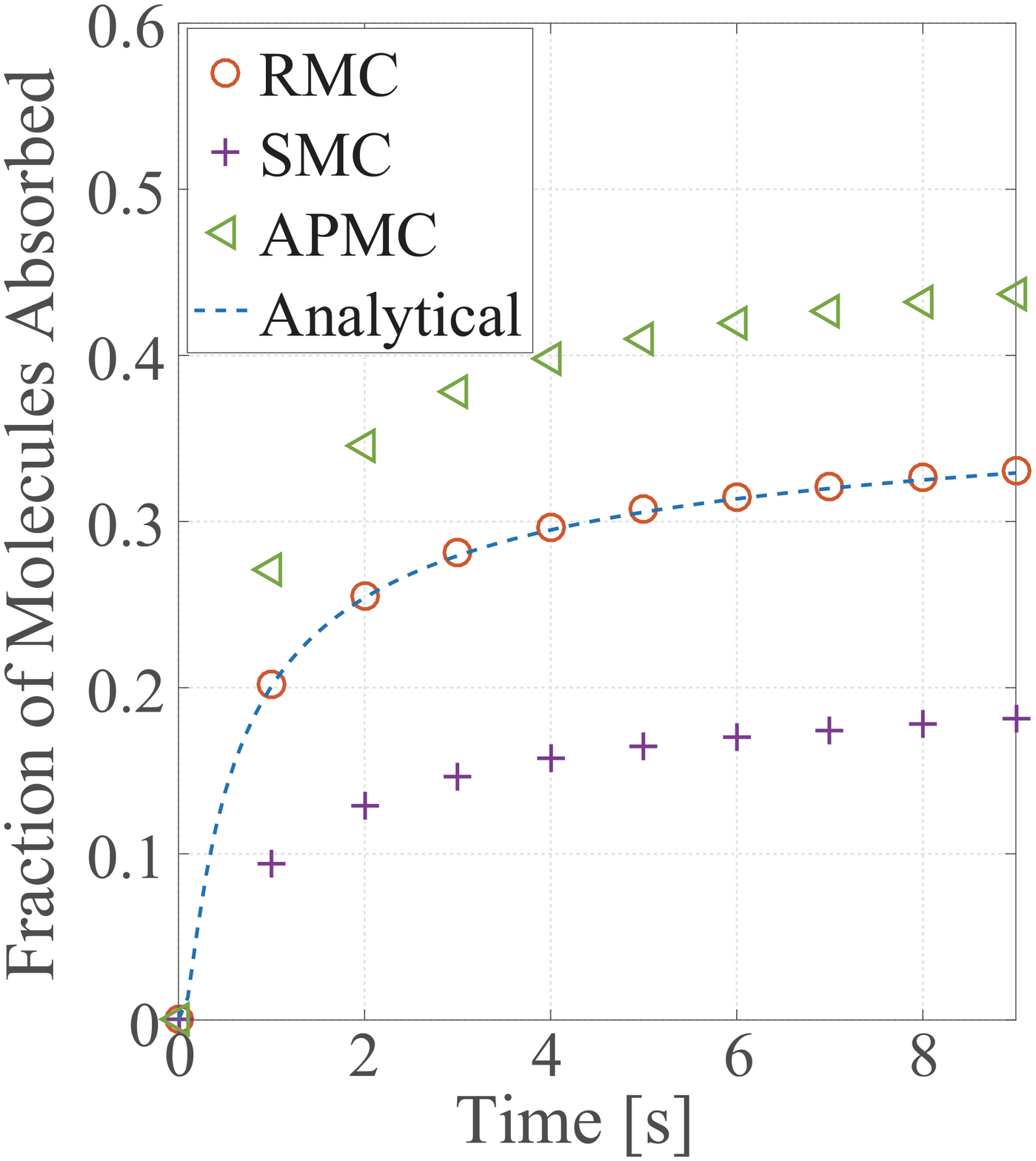}
    \caption{$\radR = 20\,\mu\m$}\end{subfigure}
    \begin{subfigure}[b]{0.24\textwidth}
    \includegraphics[width=\textwidth,height=2in]{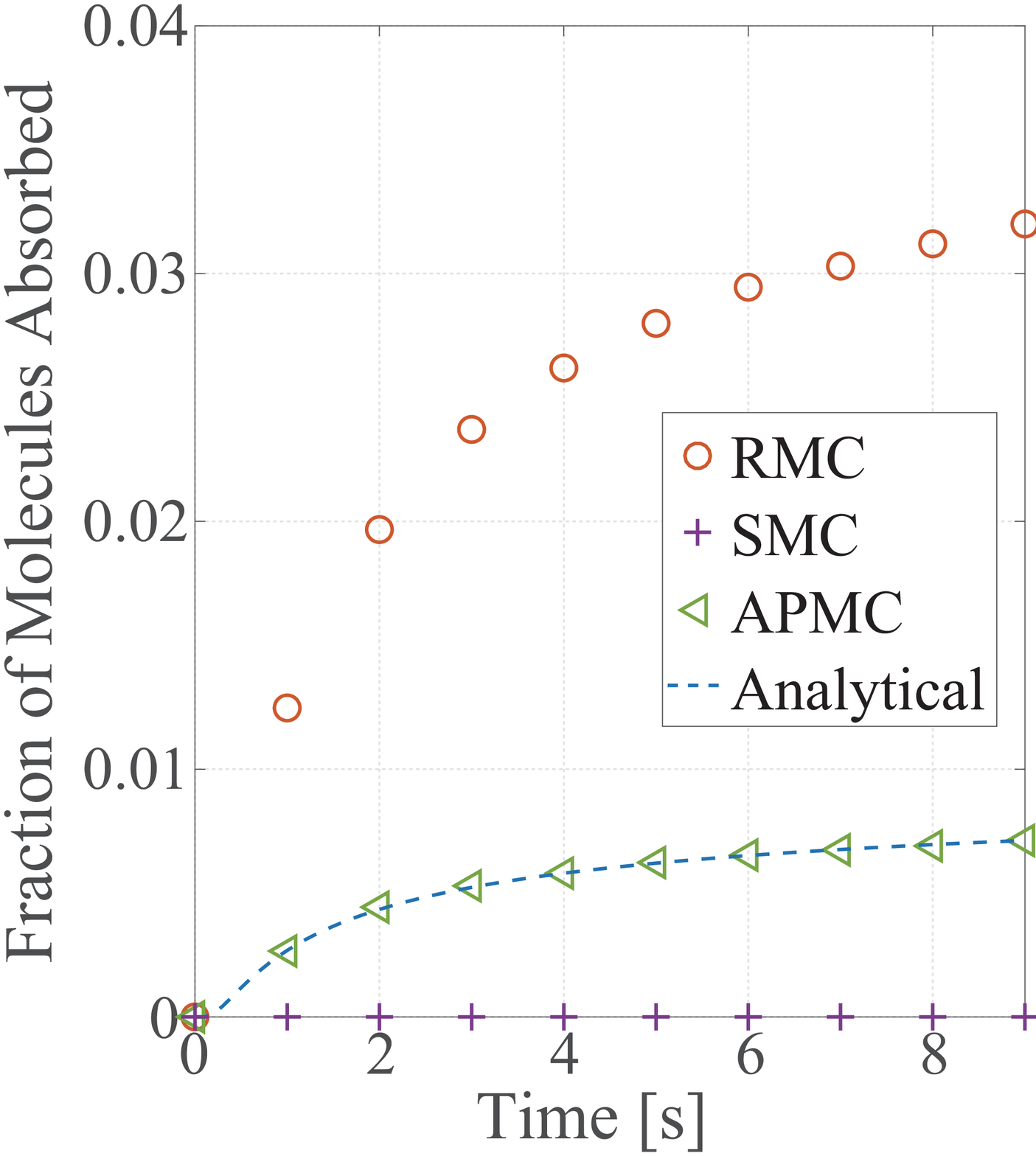}
    \caption{$\radR = 0.5\,\mu\m$}\end{subfigure}
    \caption{Comparison of the fraction of molecules absorbed results produced by the SMC algorithm, the RMC algorithm, and the APMC algorithm versus time for different $\radR$ with $M = 100$ and $\Delta t = 0.1\,\s$.\vspace{0em}}\label{timevarying1}
\end{figure}

\begin{figure}[!t]
\centering
    \begin{subfigure}[b]{0.24\textwidth}\label{planar_0106}
    \includegraphics[width=\textwidth,height=2in]{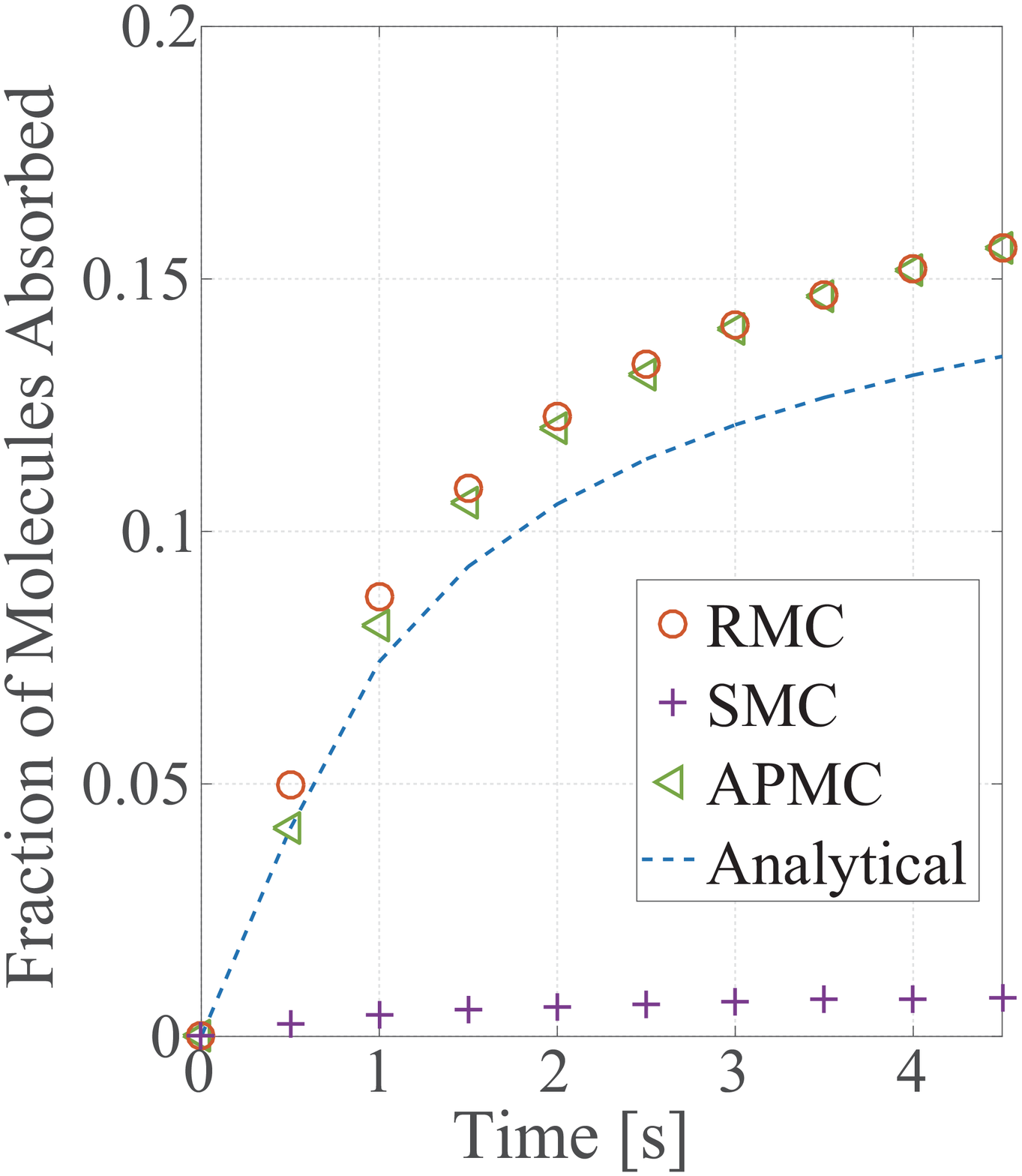}
    \caption{$\Delta t = 0.5\,\s$}\end{subfigure}
    \begin{subfigure}[b]{0.24\textwidth}\label{planar_0121}
    \includegraphics[width=\textwidth,height=2in]{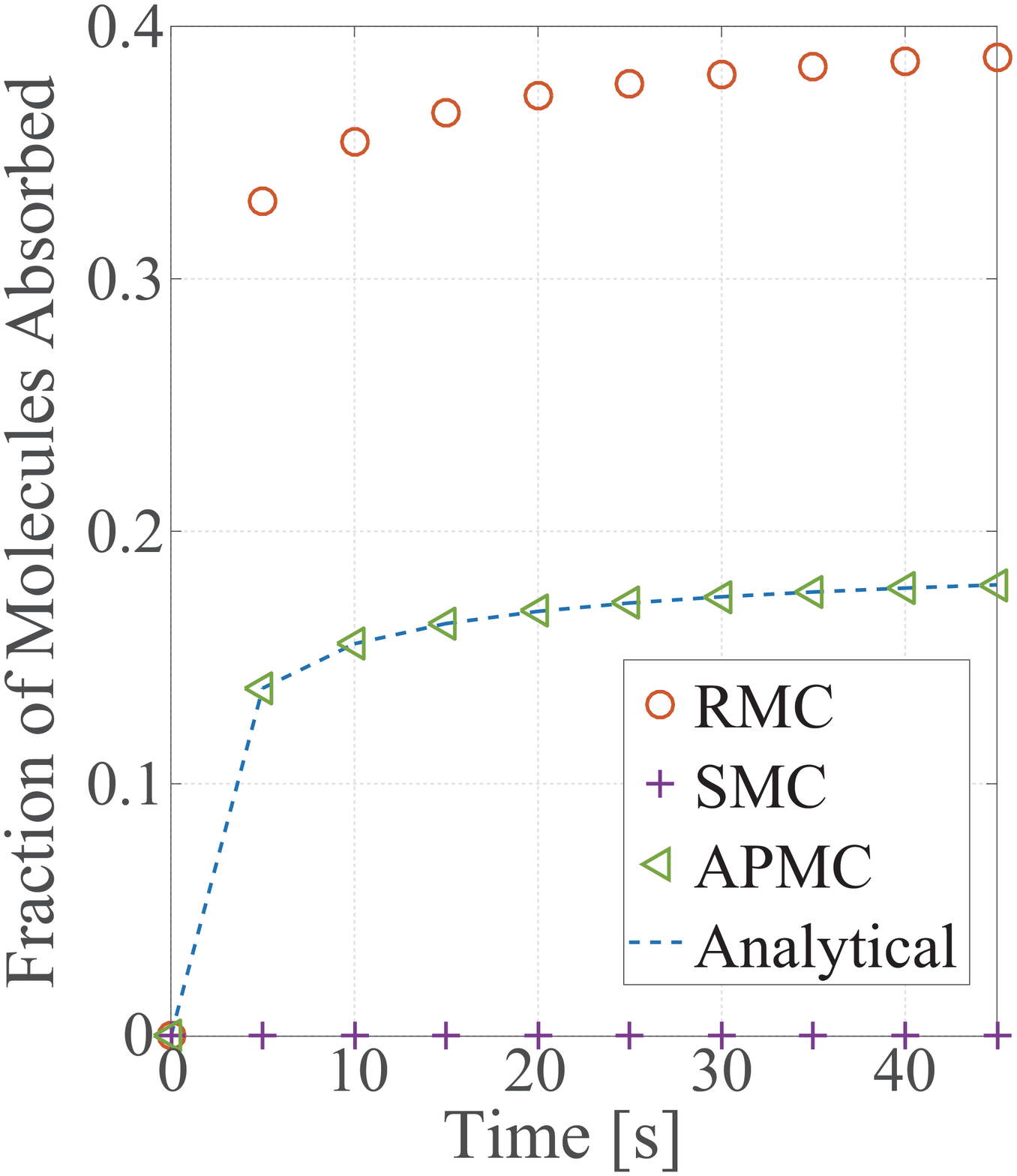}
    \caption{$\Delta t = 5\,\s$}\end{subfigure}
    \caption{Comparison of the fraction of molecules absorbed results produced by the SMC algorithm, the RMC algorithm, and the APMC algorithm versus time for different $\Delta t$ with $M = 10$ and $\radR = 10\,\mu\m$.\vspace{0em}}\label{timevarying2}
\end{figure}

In this subsection, we focus on the MC system with a single absorbing RX. We first examine the impact of $\radR$ and $\Delta t$ on the fraction of molecules absorbed produced by the three algorithms in Fig.~\ref{timevarying1} and Fig.~\ref{timevarying2}, respectively.

Fig.~\ref{timevarying1} plots the simulated fraction of molecules absorbed versus time $t$ for $M = 100$, $\Delta t = 0.1\,\s$, and $\radR = 20\,\mu \m$ or $0.5\,\mu\m$. In this figure, the analytical result obtained using \eqref{yilmazequation} is also plotted for examining the accuracy of the algorithms. From this figure, we observe that our APMC algorithm achieves a higher accuracy when $\radR$ decreases while $\Delta t$, $D$, and $\radD$ remain unchanged. Specifically, Fig.~\ref{timevarying1}(a) shows that the RMC algorithm matches the fraction of molecules absorbed as predicted by the analytical result when $\sqrt{D \Delta t}/\radR$ is small, which meets our expectation. Indeed, the performance of the RMC algorithm depends on the value of $\sqrt{D \Delta t}/\radR$. When $\sqrt{D \Delta t}/\radR$ approaches 0 and $\radD$ is larger than $\radR$, the surface area of the RX can be approximated by an infinite plane. Therefore, the probability of a molecule entering the RX between time steps is comparable to the probability of a molecule crossing a flat planar boundary between time steps. We also observe from Fig.~\ref{timevarying1}(a) that when $\sqrt{D \Delta t}/\radR$ is small, the SMC algorithm and the APMC algorithm underestimates and overestimates the fraction of molecules absorbed, respectively. Unlike Fig.~\ref{timevarying1}(a), Fig.~\ref{timevarying1}(b) shows that the APMC algorithm matches the fraction of molecules absorbed predicted by the analytical result while the other two algorithms do not. Particularly, the fractions of molecules absorbed produced by the RMC and SMC algorithms are very far away from the analytical result when $t>0\,\s$. Indeed, when $\sqrt{D \Delta t}/\radR$ is large, the spherical RX's boundary cannot be approximated by a flat planar boundary. Therefore, the RMC algorithm overestimates the fraction of molecules absorbed.

Fig.~\ref{timevarying2} plots the simulated fraction of molecules absorbed, together with the analytical result obtained using \eqref{yilmazequation}, versus time $t$ for $M = 10$, $\radR = 10\,\mu\m$, and $\Delta t = 0.5\,\s$ or $5\,\s$. We observe from this figure that when $\Delta t$ increases from $0.5\,\s$ to $5\,\s$, our APMC algorithm achieves a higher accuracy. Specifically, Fig.~\ref{timevarying2}(a) shows that the gap between the fraction of molecules absorbed produced by the APMC algorithm and that produced by the RMC algorithm is very small. Also, Fig.~\ref{timevarying2}(a) shows that both the APMC and RMC algorithms overestimate the fraction of molecules absorbed. Unlike Fig.~\ref{timevarying2}(a), Fig.~\ref{timevarying2}(b) shows that the APMC algorithm matches the fraction of molecules absorbed predicted by the analytical result. The fraction of molecules absorbed produced by the RMC algorithm is approximately twice that produced by the APMC algorithm when $t\geq10\,\s$ in Fig.~\ref{timevarying2}(b). This demonstrates the accuracy of our APMC algorithm when $\sqrt{D \Delta t}/\radR$ is large.
\begin{figure}[t]
\centering
    \begin{subfigure}[b]{0.24\textwidth}
    \includegraphics[width=\textwidth,height=2in]{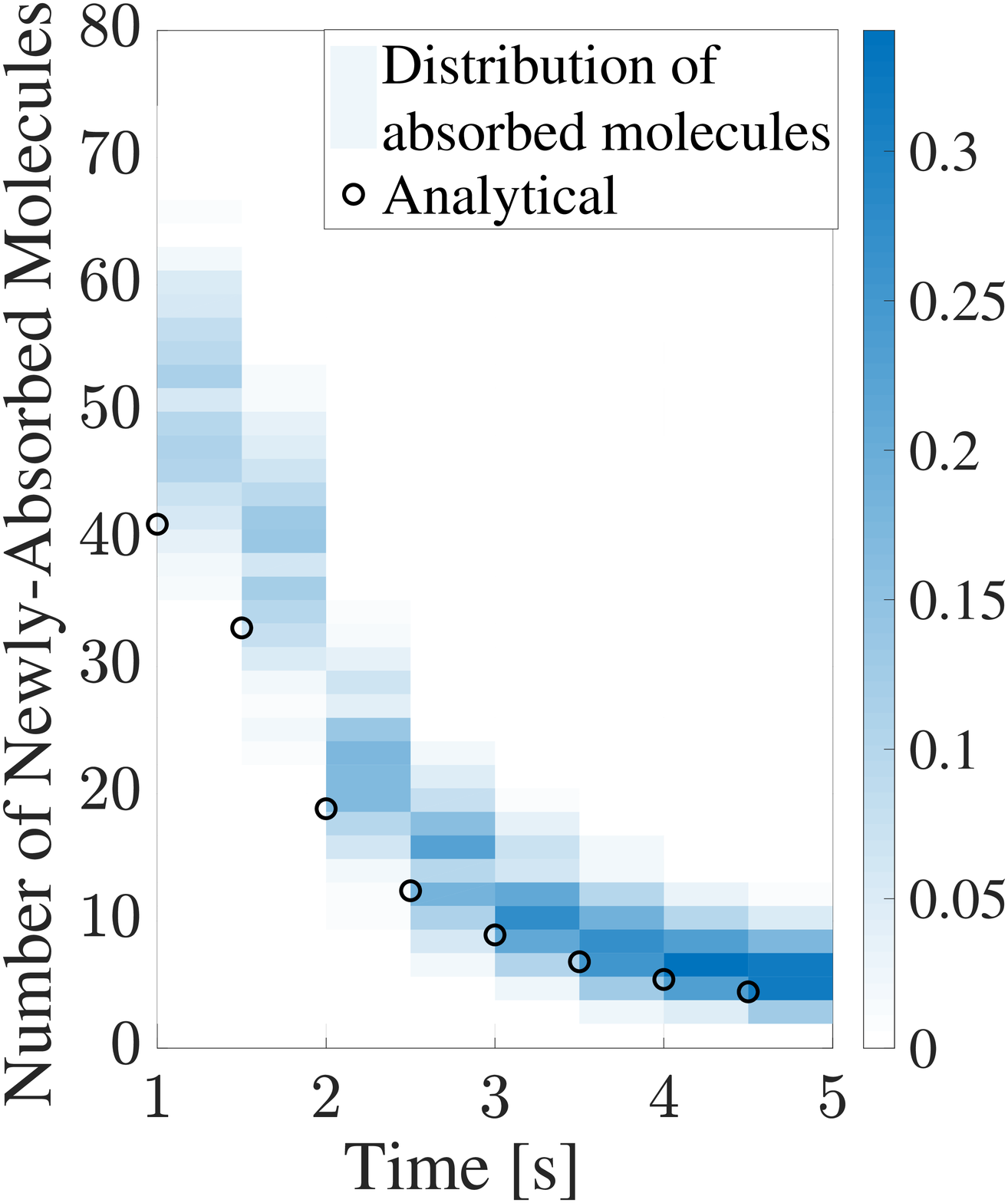}
    \caption{RMC Algorithm}\end{subfigure}
    \begin{subfigure}[b]{0.24\textwidth}
    \includegraphics[width=\textwidth,height=2in]{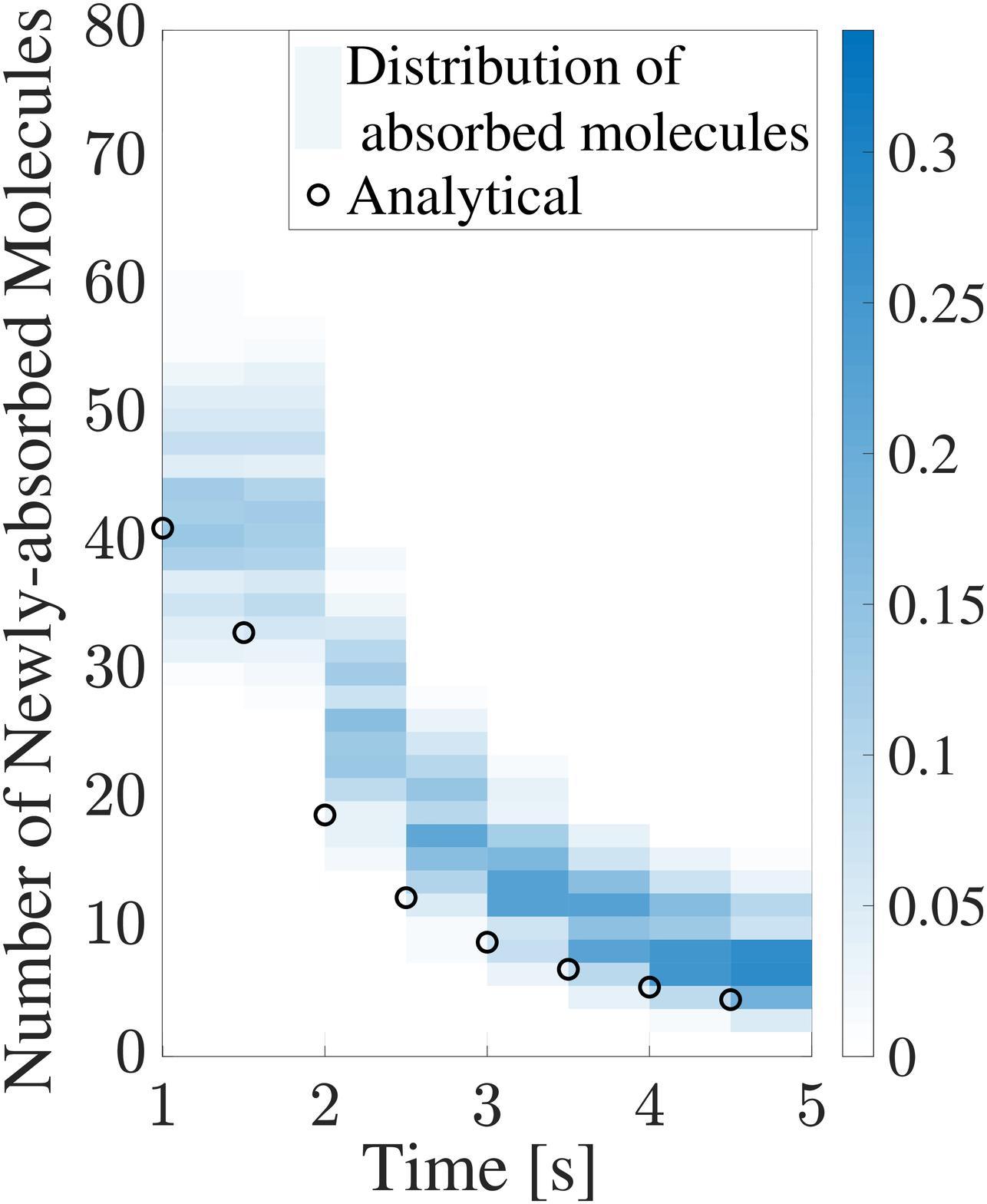}
    \caption{APMC Algorithm}\end{subfigure}
    \caption{Distribution of the number of newly-absorbed molecules during each time step for the RMC and APMC algorithms versus time for $\Delta t=0.5\,\s$, $M=10$, and $\radR=10\,\mu\m$. Simulations are repeated $10^{3}$ times and $N=10^{3}$ molecules are released each time. The spectrum bars represent observation probabilities.\vspace{-1em}}\label{Distribution}
\end{figure}

Fig.~\ref{Distribution} depicts the distribution of newly-absorbed molecules during each time step (e.g., from $t=1\,\s$ to $t=1.5\,\s$) for both the RMC and APMC algorithms versus time $t$ for $M = 10$, $\radR=10\,\mu\m$, and $\Delta t=0.5\,\s$. The analytical result in this figure is obtained using \eqref{yilmazequation}. Comparing Fig.~\ref{Distribution}(a) with Fig.~\ref{Distribution}(b), we observe that the RMC algorithm significantly overestimates the number of newly-absorbed molecules when $t=1\,\s$, while our APMC algorithm gives an improved accuracy when $t=1\,\s$ by absorbing molecules according to \eqref{analysisequationtimestep}. When $t$ increases, the overestimation of the RMC algorithm becomes slightly less severe than that of the APMC algorithm, this is in accordance with the observation from Fig.~\ref{timevarying2}(a). More importantly, we observe that the distribution of newly-absorbed molecules in Fig.~\ref{Distribution}(b) is very similar to that in Fig.~\ref{Distribution}(a), which demonstrates that our APMC algorithm does not noticeably disrupt the statistical distribution.

\begin{figure}[t]
  \centering
  \begin{subfigure}{0.24\textwidth}
  \includegraphics[width=\textwidth,height=2in]{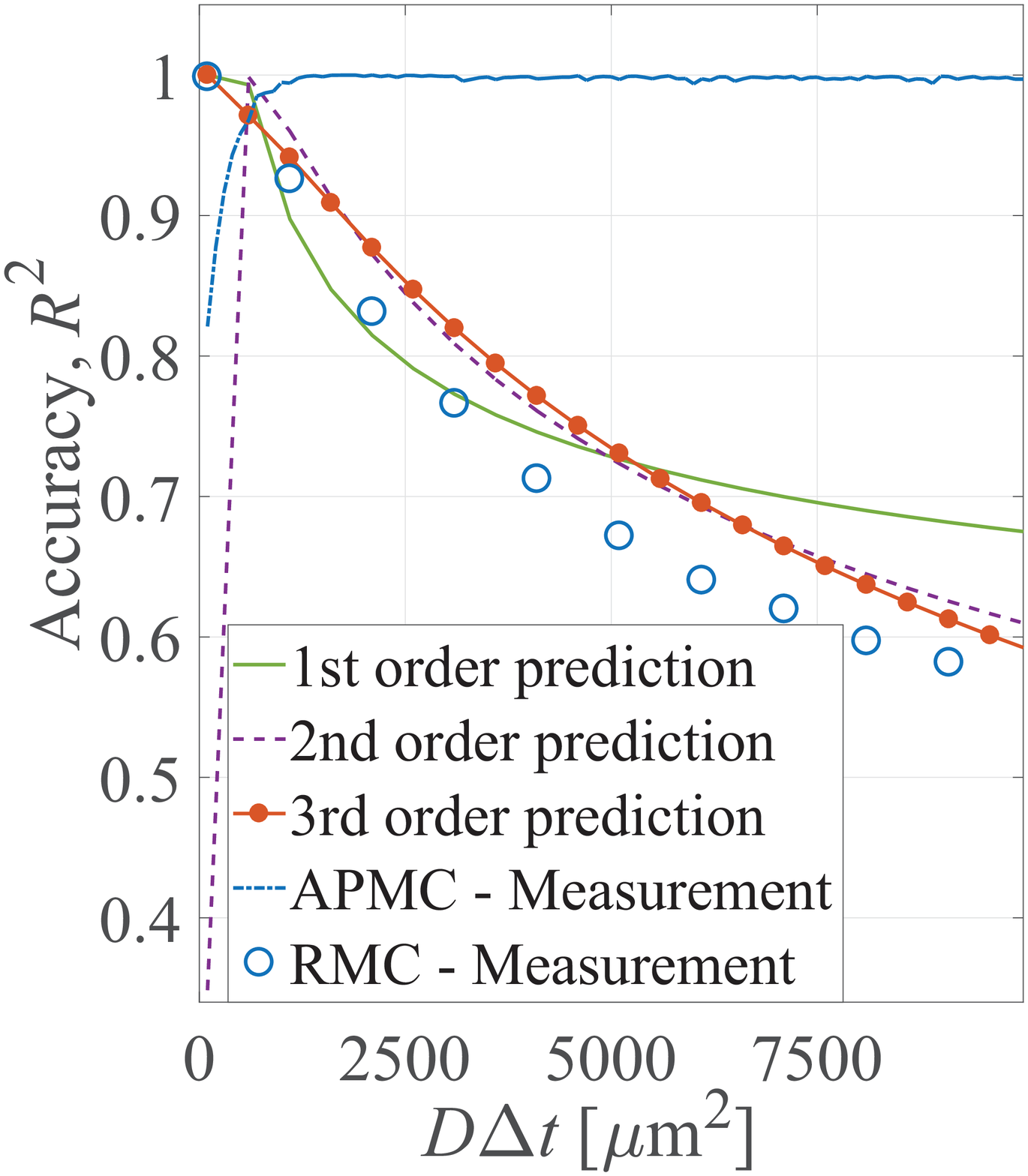}
      \caption{$\radR = 15\,\mu\m$}\end{subfigure}
       \begin{subfigure}{0.24\textwidth}
  \includegraphics[width=\textwidth,height=2in]{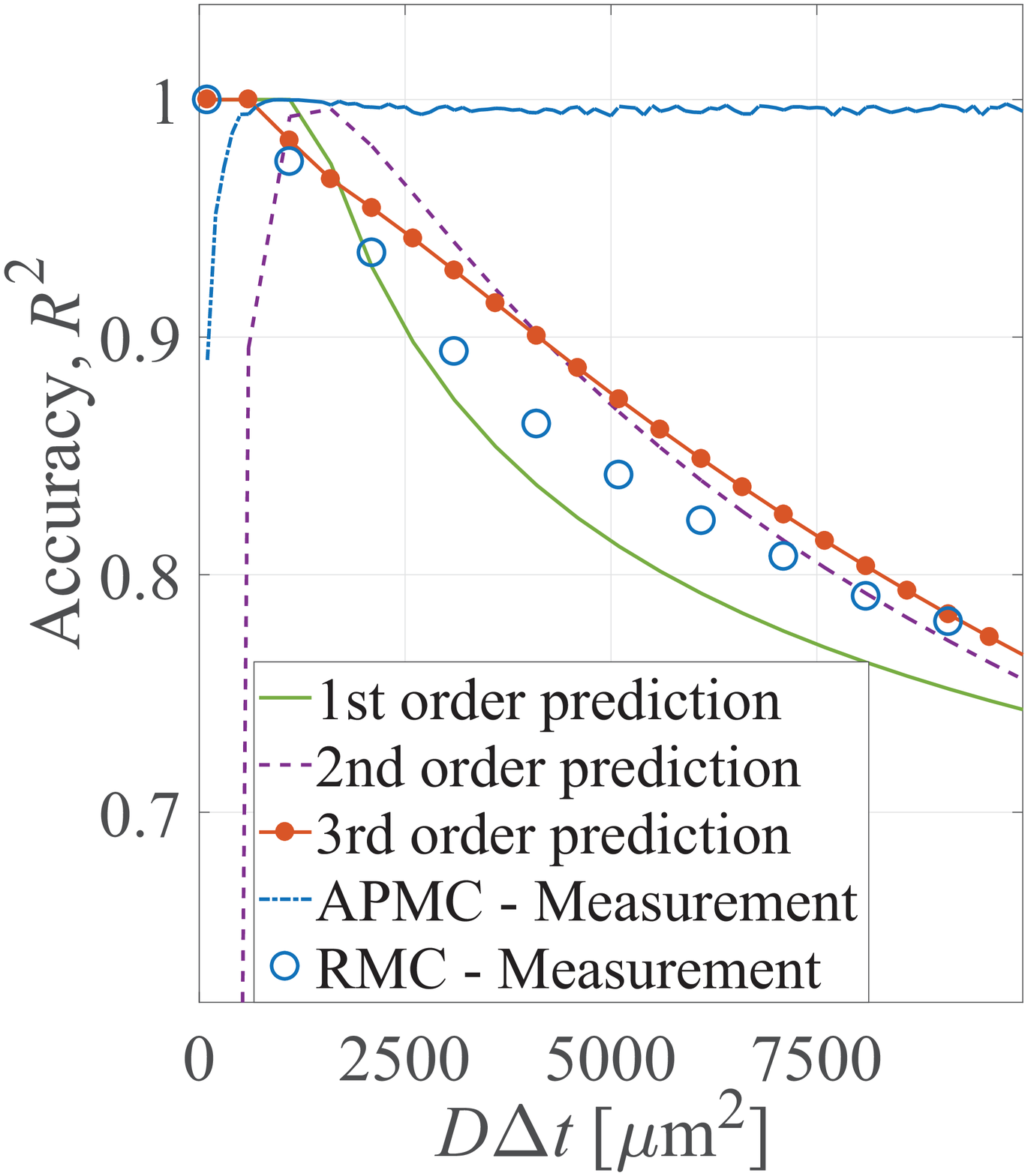}
      \caption{$\radR = 20\,\mu\m$}\end{subfigure}
   \caption{Measured and predicted accuracy of the RMC algorithm and the measured accuracy of the APMC algorithm versus $D \Delta t$ for different $\radR$ with $\radD=40\,\mu\m$ and $M = 2$.}\label{estimation}
\end{figure}
\begin{table}[t]
\renewcommand{\arraystretch}{1.2}
\centering
\caption{RMSE Measurements for Polynomial Fits in Fig.~\ref{estimation}}\label{tableRMSE}
\begin{tabular}{c||c|c|c}
\hline
 & $\textbf{1st order fit}$ & $\textbf{2nd order fit}$ & $\textbf{3rd order fit}$ \\\hline
\hline
Fig.~\ref{estimation}(a) & 0.0638 & 0.1487 & 0.0440\\\hline
Fig.~\ref{estimation}(b) & 0.0255 & 0.4378 & 0.0218\\\hline
\end{tabular}
\end{table}
We now examine the predicted accuracy of the RMC algorithm by comparing it with the measured one. The expressions for the first, second, and third order polynomial fits to predict accuracy are given by \eqref{1storder}, \eqref{2storder}, and \eqref{performance2}, respectively. We note that the calculation of \eqref{1storder}, \eqref{2storder}, and \eqref{performance2} requires $\kappa$ only. As given by \eqref{estimationeq}, $\kappa$ is a function of $D$, $\Delta t$, $\radD$, and $\radR$.

Fig.~\ref{estimation} plots the predicted accuracy given by \eqref{1storder}, \eqref{2storder}, and \eqref{performance2} as well as the measured accuracy of the APMC and RMC algorithms versus $D\Delta t$ for $\radD = 40\,\mu \m$, $M=2$, and $\radR = 15\,\mu \m$ or $20\,\mu\m$ . We observe from Fig.~\ref{estimation} that when $D\Delta t$ increases, for a fixed $\radR$, the accuracy of the RMC algorithm is at first higher than that of the APMC algorithm, but soon becomes much lower than that of the APMC algorithm. We also observe that the accuracy of the APMC stays close to 1 for most of the range of $D\Delta t$ considered. This demonstrates the accuracy and robustness of our APMC algorithm when $\sqrt{D \Delta t}/\radR$ is large.

\begin{figure}[t]
    \centering
    \includegraphics[width=0.9\columnwidth,height=2in]{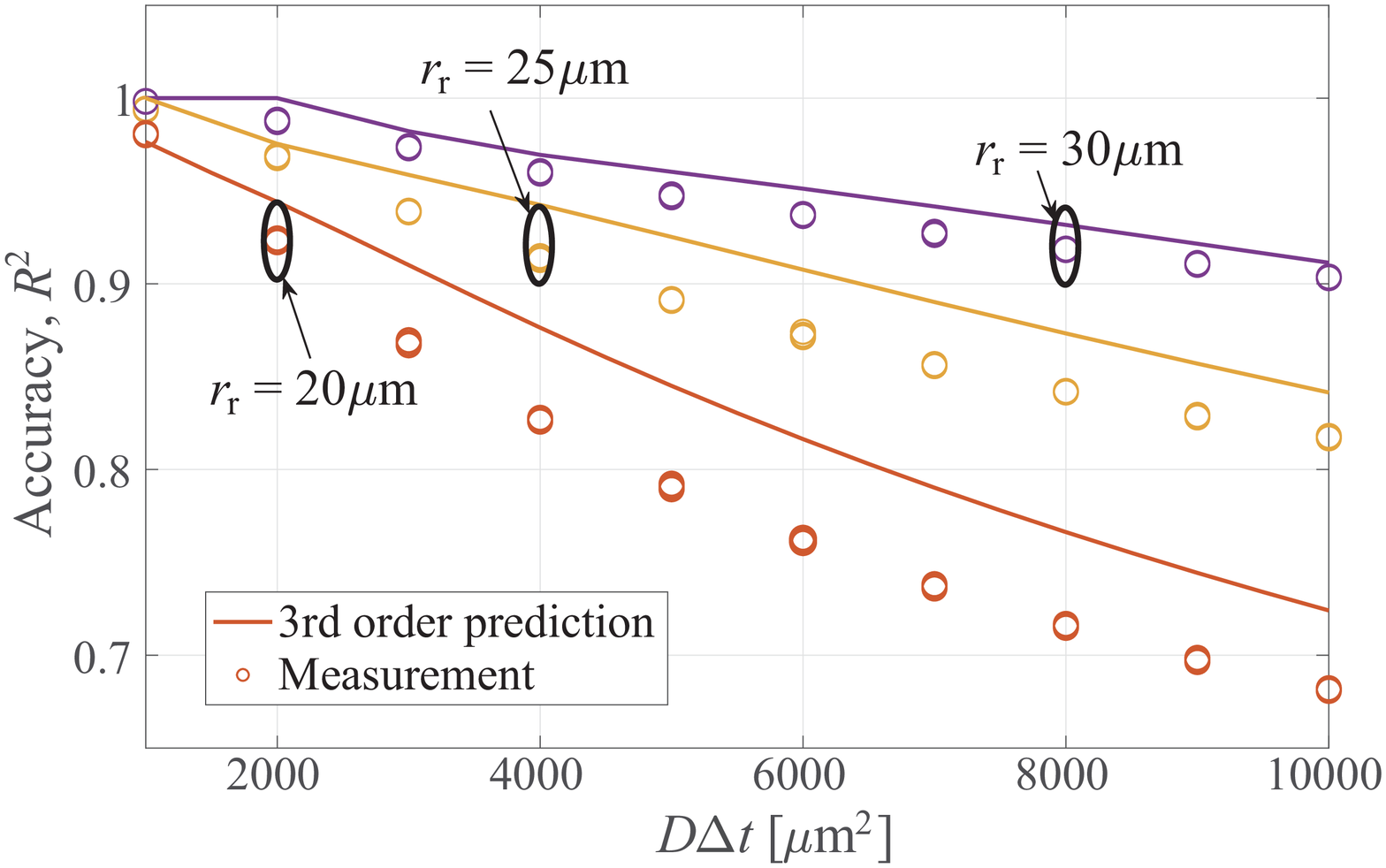}
    \caption{Measured and predicted accuracy of the RMC algorithm versus $D \Delta t$ for different $\radR$ with $M = 2$.}\label{12}
\end{figure}

\begin{figure}[t]
    \centering
    \includegraphics[width=0.9\columnwidth,height=2in]{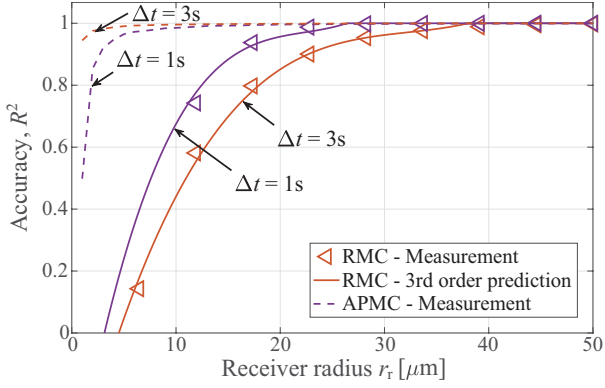}
    \caption{Measured and predicted accuracy of the RMC algorithm and the measured accuracy of the APMC algorithm versus $\radR$ for different $\Delta t$ with $M = 2$ and $\radD = 80\,\mu \m$.}\label{13}
\end{figure}

To quantitatively assess the accuracy of the first, second, and third order polynomial fits in Fig.~\ref{estimation}, we calculate the RMSE measurements of these fits via
\begin{equation}\label{RMSE}
\textrm{RMSE}=\sqrt{\frac{\sum_{i=1}^{2}\left({\textnormal{Pr}}_{\hit}(i-1) - {{\textnormal{Pr}}}_{\textnormal{sim}}(i-1)\right)^2}{2}},
\end{equation}
where $\textnormal{Pr}_{\hit}(i-1)$, $ {\textnormal{Pr}}_{\textnormal{sim}}(i-1)$, and $i$ are defined the same as in \eqref{R}.
The RMSE measurements for both subfigures are given in Table \ref{tableRMSE}. Based on both subfigures and Table \ref{tableRMSE}, we find that the third order polynomial fit for $R^2$ is more accurate than the first and second order polynomial fits for $\radR = 15\,\mu \m$ and $20\,\mu\m$, since the third order polynomial fit is, on average, closer to the measured accuracy than the first and second order polynomial fits, as shown in both subfigures, and achieves the lowest RMSE fitting measurement, as shown in Table \ref{tableRMSE}. Therefore, in Figs.~\ref{12} and \ref{13} we only consider the third order polynomial fit for the RMC algorithm.

Fig.~\ref{12} plots the measured accuracy and the predicted accuracy of the RMC algorithm versus $D \Delta t$ for $\radR = 20\,\mu \m$, $25\,\mu\m$, and $30\,\mu\m$. For the measured accuracy at a given $D\Delta t$, we run simulations for different combinations of $D$ and $\Delta t$ which achieve the same $D \Delta t$, e.g., the combination of $D=2\times10^{-9}\,{\m^2}/{\s}$ and $\Delta t=0.5\,\s$ or the combination of $D=1\times10^{-9}\,{\m^2}/{\s}$ and $\Delta t=1\,\s$. The simulated result for each combination is displayed by a point (i.e., circle) in a scatter plot. We observe that the points overlap with each other for a given $D\Delta t$, which demonstrates that the performance of the RMC algorithm depends on $D\Delta t$, but not $D$ or $\Delta t$ separately. We further find that the average absolute value of the difference between the measured accuracy and the predicted accuracy is $5.66\%$ for $\radR = 20\,\mu \m$, $3.27\%$ for $\radR = 25\,\mu \m$, and $1.45\%$ for $\radR = 30\,\mu \m$, all of which are lower than $6\%$.

Fig.~\ref{13} plots the measured and predicted accuracy of the RMC algorithm together with the measured accuracy of the APMC algorithm versus $\radR$ for large time step lengths $\Delta t = 1\,\s$ and $3\,\s$. As we observe from this figure, the accuracy of the APMC algorithm stays close to 1 for most of the range of $\radR$ considered. This demonstrates that for large time step lengths, the APMC algorithm preserves a very high accuracy, which agrees with the discussion on Fig.~\ref{timevarying2}(b), and the variance in accuracy is small. In addition, we observe that the measured accuracy of the RMC algorithm agrees well with the third order polynomial fit given in \eqref{performance2}. This is in accordance with the observation made from Fig.~\ref{12} that the third order polynomial fit well approximates the measured accuracy when $D\Delta t\leq3000$.

\begin{figure}[t]
    \centering
    \includegraphics[width=\columnwidth,height=2in]{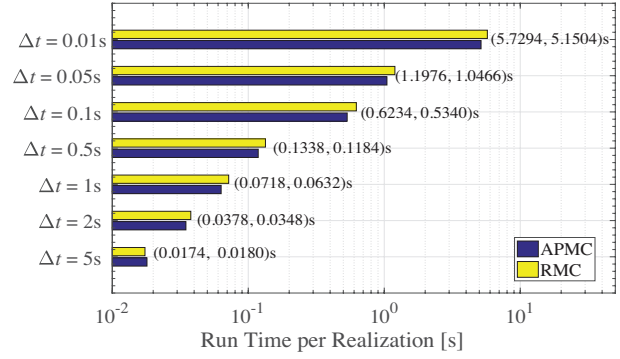}
    \caption{Average run time per realization of MATLAB simulation for the APMC and RMC algorithms for $\radR = 20\,\mu\,\m$, $N = 10^4$, and $\Delta t = 0.01\,\s$, $0.05\,\s$, $0.1\,\s$, $0.5\,\s$, $1\,\s$, $2\,\s$, and $5\,\s$.}\label{runtime}
\end{figure}

Finally, we compare the computational complexity of the APMC algorithm with that of the RMC algorithm. Fig.~\ref{runtime} plots the average run time per realization of MATLAB simulation for the APMC and RMC algorithms for $\Delta t = 0.01\,\s$, $0.05\,\s$, $0.1\,\s$, $0.5\,\s$, $1\,\s$, $2\,\s$, and $5\,\s$. We clarify that each run time is averaged over at least 20 realizations on an Intel i7 desktop PC. We first observe from the figure that the averaged run time per realization decreases when $\Delta t$ becomes higher. This is due to the fact that when $\Delta t$ increases, the number of time steps in the simulation decreases. This leads to a decrease in the number of propagation and absorption operations, which affects the computational complexity. Second, we observe that the APMC algorithm requires a slightly shorter run time to simulate a system transmission process than the RMC algorithm, except for $\Delta t = 5\,\s$. This is due to the fact that the APMC algorithm overestimates the number of absorbed molecules when $\sqrt{D\Delta t}/\radR$ is small. An overestimation means that more molecules are absorbed and removed permanently from the environment than expected, which leads to fewer molecules to be propagated and tracked by the system and thus reducing the computational complexity of the simulation. 
When the time step length increases to $\Delta t=5\,\s$, $\sqrt{D\Delta t}/\radR$ becomes large and the RMC algorithm leads to a more severe overestimation than the APMC algorithm. Therefore, the RMC algorithm requires a shorter time to simulate than the APMC algorithm.
\begin{figure}[t]
  \centering
  \begin{subfigure}{0.241\textwidth}
  \includegraphics[width=\linewidth]{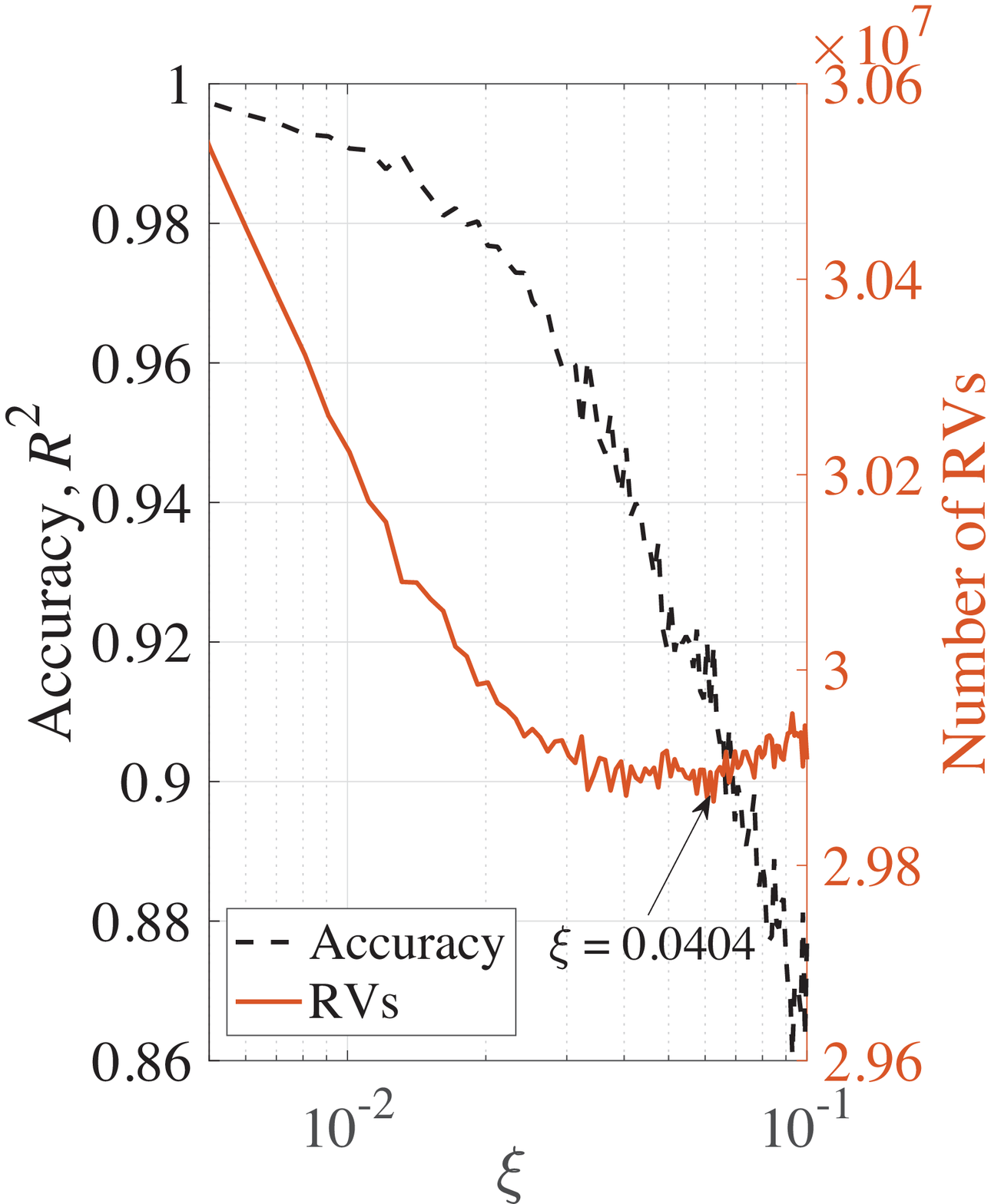}
        \caption{$\radR = 10\,\mu\m$}
        \end{subfigure}
  \begin{subfigure}{0.241\textwidth}
  \includegraphics[width=\linewidth]{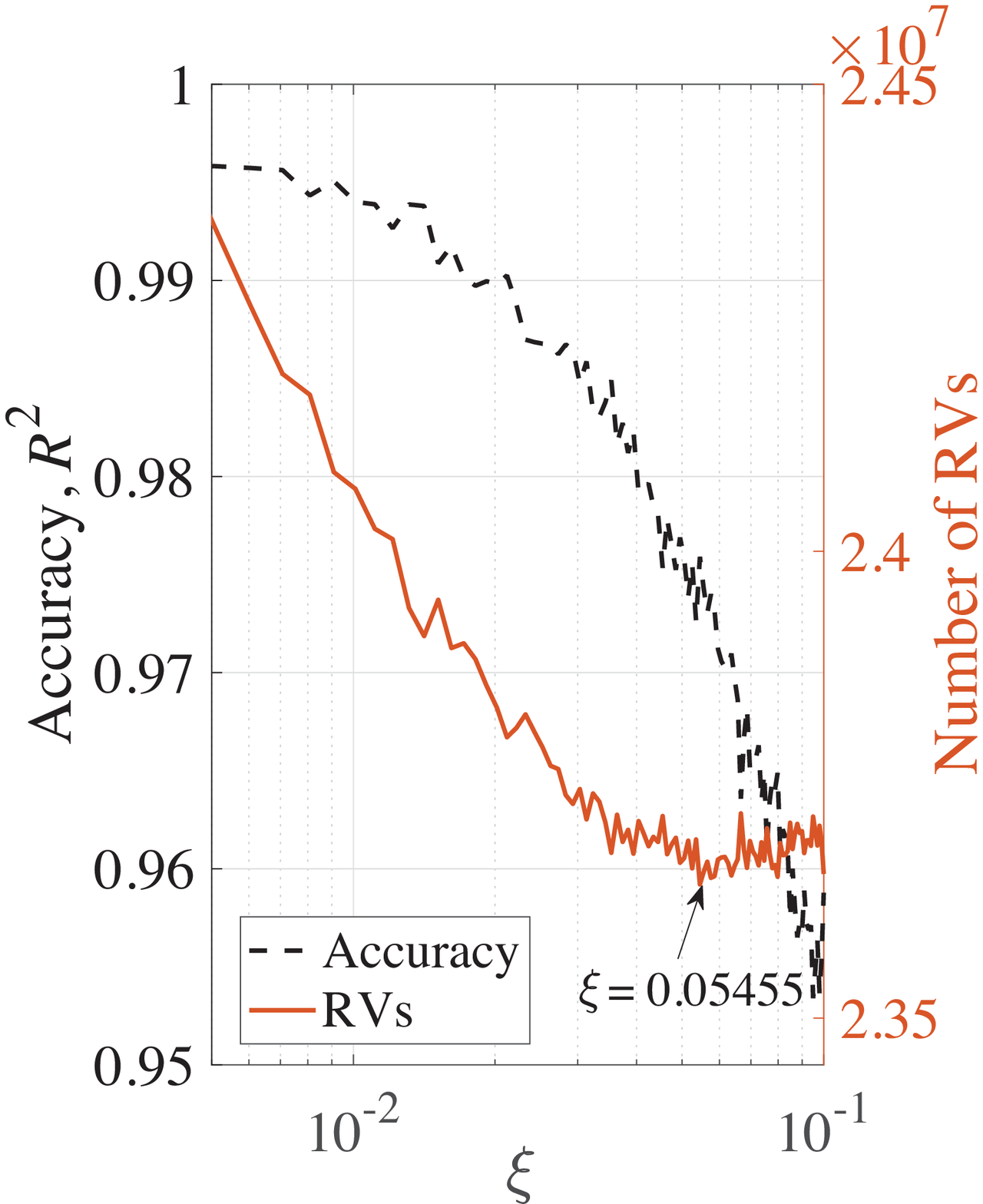}
        \caption{$\radR = 20\,\mu\m$}
        \end{subfigure}
      \caption{Number of RVs generated and measured accuracy of the APMC algorithm versus $\xi$ for $\radD = 50\,\mu\m$, $M = 10$, $N = 10^6$, $D = 10^{-9}\,{\m^2}/{\s}$, $\Delta t = 10\,\s$, and $\radR = 10\,\mu\m$ or $\radR = 20\,\mu\m$. }\label{computational_complexity_1}
\end{figure}
\begin{figure}[t]
  \centering
  \begin{subfigure}{0.241\textwidth}
  \includegraphics[width=\linewidth]{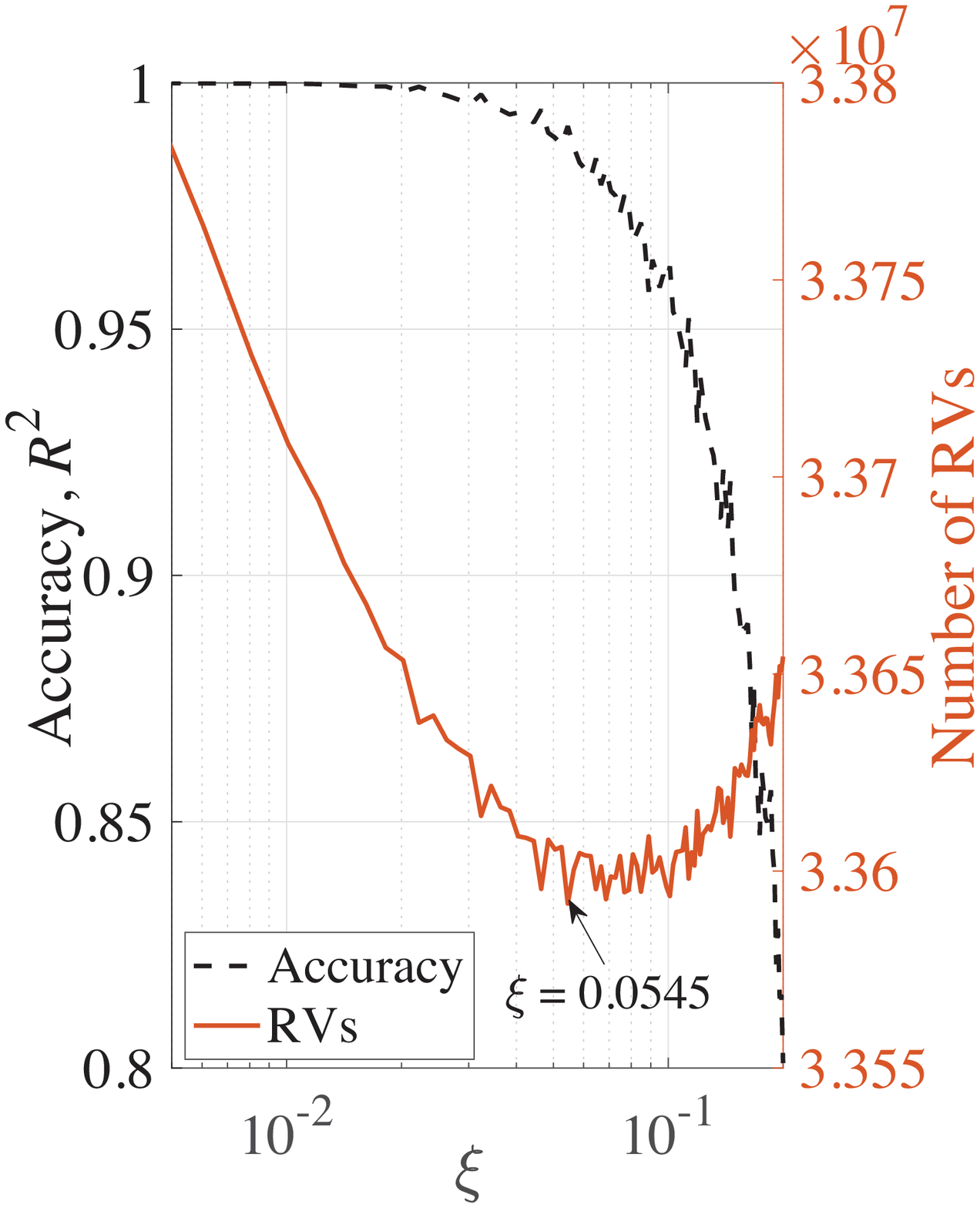}
        \caption{$\radR = 10\,\mu\m$}\end{subfigure}
    \begin{subfigure}{0.241\textwidth}
  \includegraphics[width=\linewidth]{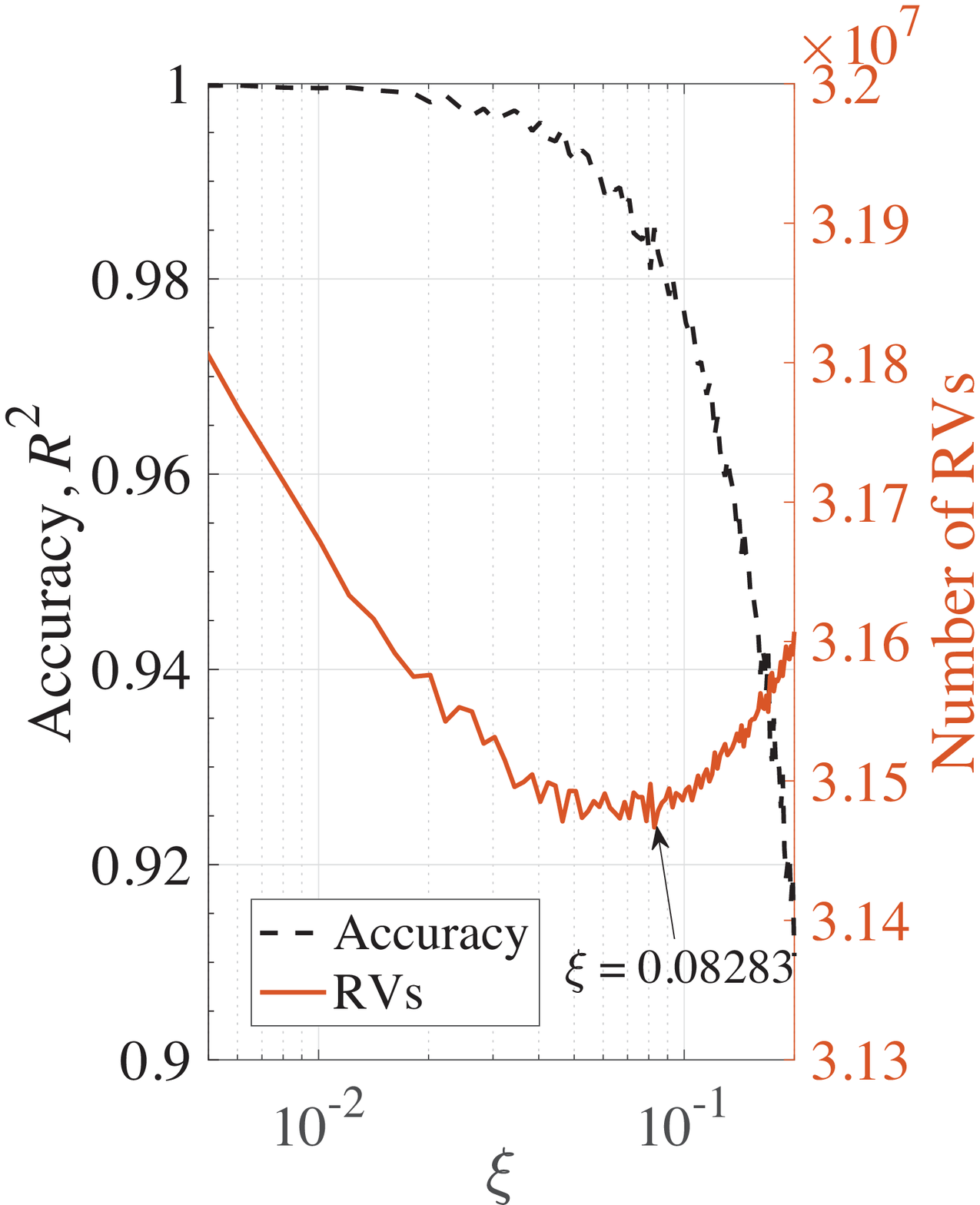}
        \caption{$\radR = 20\,\mu\m$}\end{subfigure}
   \caption{Number of RVs generated and measured accuracy of the RMC algorithm versus $\xi$ for $\radD = 50\,\mu\m$, $M = 10$, $N = 10^6$, $D = 10^{-9}\,{\m^2}/{\s}$, $\Delta t = 0.1\,\s$, and $\radR = 10\,\mu\m$ or $\radR = 20\,\mu\m$. }\label{computational_complexity}
\end{figure}

Figs.~\ref{computational_complexity_1} and \ref{computational_complexity} plot the computational complexity and the measured accuracy of the APMC algorithm and the RMC algorithm, respectively, versus the likelihood threshold $\xi$ for $\radD = 50\,\mu\m$, $M = 10$, $N = 10^6$, $D = 10^{-9}\,{\m^2}/{\s}$, and $\radR = 10\,\mu\m$ or $\radR = 20\,\mu\m$. We observe in all subfigures that when $\xi$ increases, the measured accuracy decreases. This is because when a higher likelihood threshold is applied, the number of molecules that have absorption probabilities lower than the threshold increases. 
We note that ignoring low absorption probabilities leads to an underestimation of the number of absorbed molecules, resulting in lower accuracy. It also leads to a larger number of $\mathcal{N}_\textnormal{g}$ since $\mathcal{N}_\textnormal{g}$ depends on the number of molecules that need to be propagated. We also observe that the growth in $\mathcal{N}_\textnormal{g}$ is negligible compared to the decrease in $\mathcal{N}_\textnormal{u}$ when $\xi$ is low. 
We further observe that when $\xi$ increases beyond a certain value, such as $\xi = 0.0404$ in Fig.~\ref{computational_complexity_1}(a), $\xi = 0.05455$ in Fig.~\ref{computational_complexity_1}(b), $\xi = 0.0545$ in Fig.~\ref{computational_complexity}(a), and $\xi = 0.08283$ in Fig.~\ref{computational_complexity}(b), the number of generated RVs begins to increase, since the growth in $\mathcal{N}_\textnormal{g}$ eventually outnumbers the reduction in $\mathcal{N}_\textnormal{u}$.

In addition, we observe from Fig.~\ref{computational_complexity_1} and Fig.~\ref{computational_complexity} that when $\radR$ increases from $10\,\mu\m$ to $20\,\mu\m$, the change in measured accuracy becomes less severe when applying the same $\xi$. This is because a larger RX leads to a higher \textit{a priori} absorption probability in the APMC algorithm and a higher intra-step absorption probability in the RMC algorithm. Therefore, the impact of the likelihood threshold on the number of absorptions is smaller for a larger RX. 
We clarify that Figs.~\ref{computational_complexity_1} and~\ref{computational_complexity} reveal the impact of the likelihood threshold on the number of required RVs and the measured accuracy. With the aid of these figures and similar results, a suitable likelihood threshold can be identified to achieve an acceptable loss of accuracy for simulating other system parameters. 

\subsection{Multiple Absorbing Receivers}\label{sec:Numerical_Results_Two}

\begin{figure}[t]
  \centering
  \includegraphics[width=0.88\columnwidth]{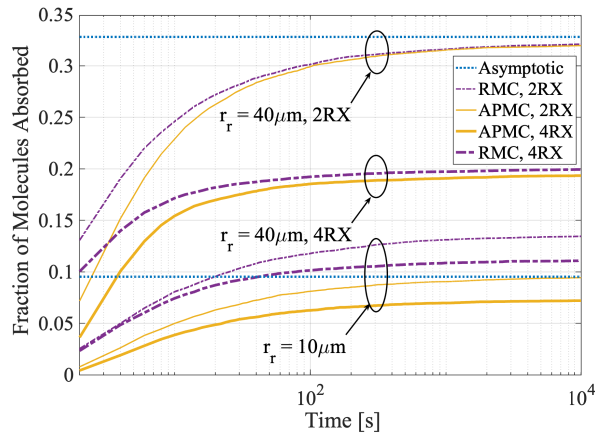}
   \caption{Comparison of the fraction of molecules absorbed for a system with two perfectly absorbing RXs and a system with four perfectly absorbing RXs. Results are produced by the RMC algorithm and the APMC algorithm versus time for $\radD = 100\,\mu\m$,  $M = 5\times10^3$, $D = 1.05\times10^{-9}\,{\m^2}/{\s}$, $\Delta t = 2\,\s$, with $\radR = 10\,\mu\m$ and $\radR = 40\,\mu\m$. The asymptotic fraction of absorbed molecules of one of the RXs in a system with two perfectly absorbing RXs as time goes to infinity is also plotted. }\label{multifinal}
\end{figure}

In this subsection, we focus on the MC system with multiple absorbing RXs. We run simulations for the system with two absorbing RXs, placed symmetrically as described in the last paragraph of Section II, and the system with four absorbing RX, where the TX is a point located at the origin, and the RXs are located at $(\radD,0,0)$, $(-\radD,0,0)$, $(0,\radD,0)$, and $(0,-\radD,0)$ in a Cartesian coordinate system with the radius as $\radR$. For both systems, we examine the fraction of molecules absorbed as produced by the RMC and APMC algorithms. Due to the symmetry of the systems, the probabilities that a molecule is absorbed by each of the RXs are identical. 

Fig.~\ref{multifinal} plots the simulated fraction of molecules absorbed by one of the RXs versus time $t$ for $\radD = 100\,\mu\m$, $M = 5\times10^3$, $\Delta t = 2\,\s$, and $\radR=\{10, 40\}\,\mu\m$ for the system with two absorbing RXs and the system with four absorbing RXs. For the system with two absorbing RXs, we compare the simulated results with the asymptotic fractions of molecules absorbed calculated using \eqref{solution} to examine the accuracy of simulations. This is due to a lack of time-varying analytical results for the fraction of molecules absorbed by two absorbing RXs in the literature. We observe that when $\radR = 10\,\mu\m$, the fraction of molecules absorbed as produced by the APMC algorithm approaches the asymptotic results when $t$ is large, while the fraction of molecules absorbed as produced by the RMC algorithm is much higher than the asymptotic results. This is because when $\radR$ is small, the spherical RX's boundary cannot be accurately approximated by a flat planar boundary. We also observe that when $\radR$ increases to $40\,\mu\m$, the fractions of molecules absorbed as produced by both algorithms approach the asymptotic results when $t$ is large. Indeed, when $\radR$ increases, the approximation of the spherical RX's boundary by a flat planar boundary becomes more accurate. For the system with four absorbing RXs, we observe that when $\radR = 10\,\mu\m$, the fraction of molecules absorbed as produced by the APMC algorithm is much lower than that as produced by the RMC algorithm. Specifically, the RMC algorithm approaches a value which is even higher than the asymptotic probability of a molecule being absorbed by one of the RXs in the two-RX system when $t$ is large, which is an intuitive overestimation. When $\radR$ increases to $40\,\mu\m$, the fractions of molecules absorbed as produced by both algorithms approach the same value, which is similar to the system with two absorbing RXs.

\section{Conclusion}\label{sec:Conclusion}

In this paper, we proposed a new algorithm named the $\apriori$ Monte Carlo (APMC) algorithm to use the $\apriori$ approach for simulating the fraction of molecules being absorbed by spherical receiver(s). Based on numerical results, we demonstrated that the APMC algorithm outperforms the RMC algorithm for large $\sqrt{D\Delta t}/\radR$. Thus, the APMC algorithm is suitable to be used in simulations for both the single absorbing receiver case and the two absorbing receivers case when $\sqrt{D\Delta t}/\radR$ is large. Moreover, for the single absorbing receiver case, we proposed an expression to predict the simulation accuracy of an existing algorithm, the refined Monte Carlo (RMC) algorithm. We further investigated the MATLAB run time of the APMC and RMC algorithms and applied likelihood thresholds to reduce the computational complexity of both algorithms. This investigation revealed that the APMC algorithm significantly reduces the computational complexity of simulating absorbing receivers in MC systems without compromising accuracy. Thus, it is worthwhile to extend the APMC algorithm to other reactive surfaces where the $\apriori$ probability of surface interaction is known.

\bibliographystyle{IEEEtran}
{\footnotesize{\bibliography{refs}}

\end{document}